\documentclass[%
reprint,
superscriptaddress,
showpacs,
preprintnumbers,
amsmath,
amssymb,
aps,
prd,
]{revtex4-2}

\usepackage{graphicx}% Include figure files
\usepackage{dcolumn}% Align table columns on decimal point
\usepackage{bm}% bold math

\begin{document}

\title{Fermion scattering on topological solitons in the nonlinear $O(3)$ $\sigma$-model}
\author{A.~Yu.~Loginov}
\email{a.yu.loginov@tusur.ru}
\affiliation{Tomsk State University of Control Systems and Radioelectronics, 634050 Tomsk, Russia}

\date{\today}% It is always \today, today,

\begin{abstract}
The scattering of  Dirac  fermions  in  the  background  fields  of topological
solitons  of the $(2+1)$-dimensional nonlinear $O(3)$ $\sigma$-model is studied
using  both analytical and numerical methods.
General formulae  describing  fermion scattering are  obtained and the symmetry
properties of the partial scattering amplitudes  and elements of the $S$-matrix
are determined.
Within the  framework  of  the  Born  approximation, the scattering amplitudes,
differential  cross-sections,  and   total  cross-sections  of  fermion-soliton
scattering  are  obtained in  analytical  forms, and  their symmetry properties
and asymptotic behavior are investigated.
The dependences of the first several partial elements of the $S$-matrix  on the
momentum  of  the  fermion  are  obtained  using  numerical  methods,  and some
properties of these dependences are ascertained and discussed.
\end{abstract}

%\pacs{11.27.+d, 11.10.Lm, 11.15.-q} %11.10.Kk
\maketitle

\section{\label{sec:I} Introduction}

Topological     solitons      of     $(2   +   1)$-dimensional   field   models
\cite{Manton, Rajaraman, Rubakov} play  an  important  role  in  field  theory,
high-energy  physics,  condensed matter physics, cosmology, and  hydrodynamics.
Of these, the  vortex solutions  of  the effective  theory of superconductivity
\cite{abr} and those of  the $\left(2+1\right)$-dimensional Abelian Higgs model
\cite{nielsen} are notable.
Another  important  example  is  provided  by  the  soliton  solutions  of  the
$(2+1)$-dimensional nonlinear $O\left(3\right)$ $\sigma$-model \cite{BP}.
The invariance of  the  static  energy  functional  of  the nonlinear $O\left(3
\right)$ $\sigma$-model under scale transformations results in  a corresponding
zero mode of the quadratic fluctuation  operator in the functional neighborhood
of a given soliton solution.
As a  consequence,  the  $\sigma$-model  possesses  a  one-parameter  family of
soliton solutions with  the  same  energy  but  different  sizes, rather than a
soliton solution of fixed size.

Derrick's  theorem  \cite{derrick} offers a  number  of  ways  to  fix the size
of the soliton.
One of these is the addition of a potential term and a fourth-order term in the
field derivatives  to   the   Lagrangian  of  the nonlinear $O\left( 3 \right)$
$\sigma$-model.
The  modified nonlinear $O\left(3\right)$ $\sigma$-model  is  known as the baby
Skyrme model, and its topological solitons \cite{Bogolubskaya_1989, psz1, psz2}
(known  as  baby Skyrmions) have a fixed size.
The   baby   Skyrme    model   and   other   modifications   to  the  nonlinear
$O\left(3\right)$ $\sigma$-model have applications  in condensed matter physics
\cite{BY_1989, sondhi_1993, B_1995, ack_2014}.
This model also has applications in cosmology; it was shown  in
                Refs.~\cite{kodama_2009, kodama_2010, delsate_2011} that in the
six-dimensional Einstein-Skyrme  model, the  baby  Skyrmion  solutions  realize
warped compactification of the extra dimensions and gravity localization on the
four-dimensional brane for the negative bulk cosmological constant.

The baby Skyrme model is a  planar analogue of the original $(3+1)$-dimensional
Skyrme  model  \cite{skyrme_1961}, which  can  be regarded  as  an approximate,
low-energy effective  theory  of  QCD that becomes exact as the number of quark
colours becomes large \cite{Witten_1983, Witten_1983_b}.
It  was  shown  in   Refs.~\cite{anw_1983, an_1984}  that  nucleons  and  their
low-lying excitations, which  are  three-quark  bound states from the viewpoint
of QCD, arise as a  topological  soliton  of  the  Skyrme  model  (known as the
Skyrmion) and its low-lying excitations, respectively, whereas pions correspond
to linearized fluctuations in the model's chiral scalar field.

Topological solitons formed from scalar fields can interact with fermion fields
via the Yukawa interaction.
This fermion-soliton  interaction  may  have  a  significant impact on both the
scalar field of the soliton and the fermion field.
In particular, it was shown in
                            Refs.~\cite{kahana_1984a, kahana_1984b, ripka_1985}
that the Yukawa interaction  between the  fermion   and  scalar  fields  of the
$(3 + 1)$-dimensional  chiral-invariant  linear  $\sigma$-model results in  the
existence of  the  chiral  soliton,  a  stable configuration of the interacting
scalar and fermion fields that possesses bound fermion states.
It was also  shown  in Refs.~\cite{gold_1981, balach_1982, balach_1983} that an
external chiral field can polarize the  fermion  vacuum;  this polarization can
be interpreted as a contribution  to  the  baryon  charge  of  a chiral soliton
\cite{kahana_1984b, ripka_1985}.
Fermion bound states also exist in  the background field of a Skyrmion, and the
properties of these bound states were investigated in
                                           Refs.~\cite{hiller_1986, zhao_1989}.
The main feature of the fermion-soliton interaction found in
Refs.~\cite{kahana_1984a, kahana_1984b, ripka_1985, hiller_1986, zhao_1989}  is
that the  spectrum  of  the  Dirac  operator  shows  a  spectral  flow  of  the
eigenvalues, where the number of zero-crossing normalized bound  modes is equal
to the winding number of the soliton.
The  planar  baby  Skyrmions  inherit   this  feature  of  the  fermion-soliton
interaction. %of the original tree-dimensional Skyrmions.
It was shown in Refs.~\cite{kodama_2009, kodama_2010, delsate_2011} that in the
background field of a baby Skyrmion, the Dirac operator also shows  a  spectral
flow of the eigenvalues, and  the  number  of  zero-crossing normalized bounded
modes turns out to be equal to the winding number of the baby Skyrmion.
This property of the Dirac operator is preserved  when  the backreaction of the
fermion  field  coupled   with   the   baby  Skyrmion  is  taken  into  account
\cite{shnir_2018}.

Except for the Dirac sea fermions, only fermion bounded states belonging to the
discrete spectrum of the Dirac operator  were  considered  in  the  works cited
above.
One main aim of the present paper is  to consider the fermion scattering states
belonging to the continuous  spectrum of the Dirac operator.
The topological  solitons  of  the  nonlinear  $O(3)$ $\sigma$-model are chosen
as background fields for the Dirac operator, since the nonlinear $O(3)$ $\sigma
$-model  is  one of the few models for which exact analytical soliton solutions
are known.
Having obtained the analytical  soliton  solution  as a background field of the
Dirac operator makes it possible to obtain  some analytical results relating to
fermion scattering.
In particular,  the  scattering  amplitudes,  differential  cross-sections, and
total  cross-sections  can  be  obtained  in  an  analytical  form  within  the
framework of the first Born approximation.

This  paper  is  structured as follows.
In Sec.~\ref{sec:II}, we describe briefly  the  Lagrangian,   symmetries, field
equations, and  topological solitons  of the  nonlinear  $O(3)$ $\sigma$-model.
In Sec.~\ref{sec:III}, general properties of fermion scattering are considered,
such as the forms of the fermion  scattering  wave  functions, their asymptotic
behavior, and the symmetry properties of the partial elements of the $S$-matrix.
In Sec.~\ref{sec:IV}, we give an analytical  description  of fermion scattering
within the framework of the first Born approximation.
In Sec.~\ref{sec:V},  we  present  numerical  results  for  the  first  several
partial elements of the $S$-matrix.
In the final section, we  briefly  summarize the results obtained in this work.
Appendix A contains some  necessary  information   about   the  plane-wave  and
cylinder-wave states of free fermions.
A resonance behavior of some partial elements of the $S$-matrix is explained in
Appendix B.

Throughout this paper, we use the natural units $c = 1$, $\hbar = 1$.

\section{\label{sec:II}  Lagrangian, field  equations, and topological solitons
of the model}

The model we are interested in  is  the  $(2 + 1)$-dimensional nonlinear $O(3)$
$\sigma$-model, which includes a fermion field.
The scalar isovector  field $\boldsymbol{\phi}$ of the model interacts with the
spinor-isospinor fermion field $\psi$ via  the  Yukawa coupling, leading to the
Lagrangian
\begin{equation}
\mathcal{L}=\frac{1}{2}\partial_{\mu}\boldsymbol{\phi}
\boldsymbol{\cdot}\!\partial^{\mu}\boldsymbol{\phi}+
\frac{\lambda}{2}\left(\boldsymbol{\phi}\boldsymbol{\cdot}\!
\boldsymbol{\phi}-H^{2}\right)
+i\bar{\psi}\gamma^{\mu}\partial_{\mu}\psi+h\mathbf{\phi}
\boldsymbol{\cdot}\bar{\psi}\boldsymbol{\tau}\psi,                 \label{II:1}
\end{equation}
where $h$ is  the  Yukawa  coupling  constant  and  $\lambda$  is  the Lagrange
multiplier, which imposes the constraint $\boldsymbol{\phi}\boldsymbol{\cdot}\!
\boldsymbol{\phi} = H^{2}$ on  the  scalar isovector field $\boldsymbol{\phi}$.
In $(2+1)$ dimensions, we shall use the following Dirac matrices:
\begin{equation}
\gamma^{0}=\sigma_{3},\; \gamma ^{1}=-i\sigma_{1},\; \gamma^{2}=-i\sigma_{2},
                                                                   \label{II:2}
\end{equation}
where $\sigma_{i}$ are the Pauli matrices.
To distinguish the Pauli  matrices  $\sigma_{k}$ acting on the spinor index $i$
of the fermion field $\psi_{i,a}$ from those acting on the isospinor index $a$,
we denote the latter as $\tau_{k}$.

In natural units, the $(2+1)$-dimensional scalar field  $\boldsymbol{\phi}$ has
dimension of $\text{mass}^{1/2}$.
We adopt the squared  parameter $H^{2}$  as  the  energy  (mass)  unit, and use
dimensionless variables:
\begin{eqnarray}
\mathbf{\phi } &\rightarrow &H\mathbf{\phi },\;\psi \rightarrow H^{2}\psi ,\;
\mathbf{x}\rightarrow H^{-2}\mathbf{x}, \nonumber \\
t &\rightarrow &H^{-2}t,\;h\rightarrow Hh,\;\lambda \rightarrow H^{4}\lambda.
                                                                   \label{II:3}
\end{eqnarray}
In these new variables, the Lagrangian (\ref{II:1}) takes the form
\begin{equation}
\mathcal{L}=\frac{1}{2}\partial_{\mu}\boldsymbol{\phi}
\boldsymbol{\cdot} \partial^{\mu}\boldsymbol{\phi}+\frac{\lambda}{2}
\left(\boldsymbol{\phi}\boldsymbol{\cdot}\!\boldsymbol{\phi}-1\right)
+ i\bar{\psi}\gamma ^{\mu} \partial _{\mu}\psi + h\boldsymbol{\phi}
\boldsymbol{\cdot} \bar{\psi }\boldsymbol{\tau }\psi.              \label{II:4}
\end{equation}
By varying the action $S = \int \mathcal{L}d^{2}xdt$ in $\boldsymbol{\phi}$ and
$\bar{\psi}$, we obtain the field equations of the model:
\begin{align}
\!\!\!\partial_{\mu}\partial^{\mu}\boldsymbol{\phi}-\boldsymbol{\phi}
\left(\boldsymbol{\phi}\!\boldsymbol{\cdot}\!\partial_{\mu}\partial^{\mu}
\boldsymbol{\phi}\right)-h\boldsymbol{\phi}\left(\boldsymbol{\phi}
\boldsymbol{\cdot}\bar{\psi}\boldsymbol{\tau}\psi \right)\!+\!h
\bar{\psi}\boldsymbol{\tau}\psi & = 0,                            \label{II:5a}
\\
i\gamma ^{\mu}\partial_{\mu}\psi+h\boldsymbol{\phi}\!\boldsymbol{\cdot}\!
\boldsymbol{\tau}\psi & = 0,                                      \label{II:5b}
\end{align}
where we use the constraint $\boldsymbol{\phi} \boldsymbol{\cdot}\!\boldsymbol{
\phi}=1$ to express  the  Lagrange  multiplier $\lambda$ in terms of the fields
$\boldsymbol{\phi}$ and $\psi$.

Model (\ref{II:1}) is  invariant  under  global $SU(2)$  transformations of the
isovector field $\boldsymbol{\phi}$ and the isospinor field $\psi$.
Its classical bosonic vacuum is an arbitrary constant scalar field $\boldsymbol{
\phi}(x) = \boldsymbol{\phi}_{\text{vac}}$, which can be taken as $\boldsymbol{
\phi}_{\text{vac}} = (0, 0, -1)$.
We see  that the  vacuum  $\boldsymbol{\phi}_{\text{vac}}$  breaks the original
$SU(2)$ symmetry group of model (\ref{II:1}) to its $U(1)$ subgroup.
Any finite  energy  field  configuration  $\boldsymbol{\phi}(x)$  must  tend to
$\boldsymbol{\phi}_{\text{vac}}$ at spatial infinity.
It follows that  the  finite  energy field configurations of model (\ref{II:1})
are split into topological classes that are  elements  of  the  homotopic group
$\pi_{2}\left(S^{2}\right) = \mathbb{Z}$,  and  are  consequently characterized
by an integer winding number $n$.

It is well known \cite{BP}  that the  nonlinear $O(3)$ $\sigma$-model possesses
a variety of topological soliton solutions.
These soliton solutions exist in each topological sector  of the model, and are
absolute minima of the energy functional. %in a given sector.
In the topological sector with a given nonzero winding number $n$, the maximally
symmetric soliton solution has the form
\begin{equation}
\boldsymbol{\phi }\left( r,\theta \right) =
\begin{pmatrix}
\sin \left( f\left( r\right) \right) \cos \left( n\theta \right)  \\
\sin \left( f\left( r\right) \right) \sin \left( n\theta \right)  \\
\cos \left( f\left( r\right) \right)
\end{pmatrix}\!,                                                   \label{II:6}
\end{equation}
where $r$ and $\theta$ are the  polar  coordinates, and the profile function is
\begin{equation}
f\left( r\right) = 2\arctan \left[ \left( r/r_{0}\right) ^{\left\vert
n\right\vert }\right]\!.                                           \label{II:7}
\end{equation}
Soliton solution (\ref{II:6}) is invariant under the cyclic subgroup  $\mathbb{
Z}_{n}$ $(\theta\rightarrow \theta +2\pi k/n,\;k=0,\ldots ,n-1)$ of the spatial
rotation group $SO(2)$.
It is also invariant under the simultaneous action  of spatial rotation through
an  angle  $\delta$  and  internal  rotation  about  the  third  isotopic  axis
$\mathbf{u}_{3} = (0,0,1)$  through an angle $-n \delta$
\begin{equation}
\boldsymbol{\phi }\left( r,\theta \right) = R_{\mathbf{u}_{3}}\left( -n\delta
\right) \boldsymbol{\phi }\left( r,\theta +\delta \right),         \label{II:8}
\end{equation}
where the rotation matrix is
\begin{equation}
R_{\mathbf{u}_{3}}\left( \varphi \right) =
\begin{pmatrix}
\cos \left( \varphi \right)  & -\sin \left( \varphi \right)  & 0 \\
\sin \left( \varphi \right)  &  \cos \left( \varphi \right)  & 0 \\
0 & 0 & 1
\end{pmatrix}\!.                                                  \label{II:10}
\end{equation}
Note  that soliton  solution (\ref{II:6}) depends  on  the  positive  parameter
$r_{0}$, which can be interpreted as the spatial size of the soliton.
However, the energy of the soliton solution
\begin{eqnarray}
E &=&2\pi \int\limits_{0}^{\infty }\mathcal{E}\left( r\right) rdr=2\pi
\int\limits_{0}^{\infty }\biggl[\frac{4n^{2}}{r^{2}}\left(\frac{r}{r_{0}}
\right) ^{2\left\vert n\right\vert }\biggr.                      \nonumber
\\
& &\biggl. \times \biggl(1+\left(\frac{r}{r_{0}}\right)^{2\left\vert
n \right\vert}\biggr)^{-2}\biggr] rdr =
4\pi \left\vert n \right\vert                                     \label{II:11}
\end{eqnarray}
does not depend on $r_{0}$.
This is because the bosonic part of the  action $S = \int \mathcal{L}d^{3}x$ of
model  (\ref{II:1})  is   invariant  under  scale  transformations  $\mathbf{r}
\rightarrow a \mathbf{r}$.

\section{\label{sec:III} Fermions in the background field of the soliton}

We consider  fermion  scattering  on  the  topological soliton of the nonlinear
$O(3)$ $\sigma$-model in the external field approximation.
In this approximation, the backreaction of a  fermion on the soliton's field is
neglected, and  the fermion-soliton scattering is described solely by the Dirac
equation (\ref{II:5b}).
We can rewrite the Dirac equation in the Hamiltonian form
\begin{equation}
i\frac{\partial \psi _{i,a}}{\partial t}=H_{i,a;\,j,b}\psi_{j,b}, \label{III:1}
\end{equation}
where the Hamiltonian
\begin{equation}
H_{i,a;\,j,b}=\alpha _{i j}^{k}\otimes \mathbb{I}_{a b}\left( -i\partial
_{k}\right) -h\beta_{i j}\otimes \phi^{c}\tau_{a b}^{c},          \label{III:2}
\end{equation}
$\mathbb{I}$  is  the $2 \times 2$   identity   matrix,  $\alpha^{1} = \gamma^{
0}\gamma^{1} = \sigma_{2}$,  $\alpha^{2} = \gamma^{0}\gamma^{2} = -\sigma_{1}$,
and $\beta = \gamma^{0} = \sigma_{3}$.
In Eqs.~(\ref{III:1}) and  (\ref{III:2}),  all  spin  and  isospin  indices are
explicitly shown, and the summation convention over repeated indices is implied.

The invariance of  soliton  solution (\ref{II:6}) under combined transformation
(\ref{II:8}) results in  the existence  of  the  conserved  generalized angular
momentum or grand spin, which we denote by $K_{3}$:
\begin{equation}
\left[H, K_{3}\right] = 0,                                        \label{III:3}
\end{equation}
where
\begin{equation}
K_{3\,i,a;\,j,b}\! = \!\mathbb{I}_{i j}\otimes\mathbb{I}_{a b}\epsilon
_{k l}x_{k}(-i\partial_{l})\!+\!\frac{1}{2}\sigma_{3\,i j}
\otimes\mathbb{I}_{a b}\!+\!\frac{n}{2}\mathbb{I}_{i j}
\mathbf{\otimes}\tau_{3\,a b}                                     \label{III:4}
\end{equation}
and the $2 \times 2$ antisymmetric matrix $\epsilon_{k l} = i \sigma_{2\,k l}$.
It follows from Eq.~(\ref{III:4}) that  the  grand  spin  $K_{3}$ is the sum of
the orbital part $L_{3}=\mathbb{I}\otimes \mathbb{I}\epsilon_{kl}x_{k}\left( -i
\partial_{l}\right)$, the spin  part $s_{3} = \left(\sigma_{3}/2 \right)\otimes
\mathbb{I}$, and the  isospin  part $n I_{3} = n\mathbb{I} \otimes \tau_{3}/2$.
None of the terms $L_{3}$, $s_{3}$, and $I_{3}$  is conserved separately in the
background field of soliton solution (\ref{II:6}).
Using Eq.~(\ref{III:4}), we can obtain the eigenfunctions $\psi^{\left(m\right)
}$ of the operator $K_{3}$.
These eigenfunctions satisfy the relation %$\psi^{\left(m\right)}$
\begin{equation}
K_{3}\psi_{i a}^{\left(m\right)}=m\psi_{i a}^{\left(m\right)}   \label{III:5}
\end{equation}
and can be written in the compact matrix form
\begin{equation}
\psi_{i a}^{\left( m \right) } = \left(
\begin{array}{cc}
c_{11}\left( r\right) e^{i l_{11} \theta } & c_{12}\left( r\right)
e^{i l_{12}\theta } \\
c_{21}\left( r\right) e^{i l_{21} \theta } & c_{22}\left( r\right)
e^{i l_{22} \theta }
\end{array}
\right)\!,                                                        \label{III:6}
\end{equation}
where the orbital quantum numbers
\begin{subequations}                                             \label{III:6a}
\begin{eqnarray}
l_{11} &=&\left( 2m-n-1\right) /2, \\
l_{12} &=&\left( 2m+n-1\right) /2, \\
l_{21} &=&\left( 2m-n+1\right) /2, \\
l_{22} &=&\left( 2m+n+1\right) /2,
\end{eqnarray}
\end{subequations}
and $c_{i a}\left(r\right)$ are some radial wave functions.
The  eigenfunctions  $\psi _{i a}^{\left(  m  \right) }(r, \theta)$  should  be
single-valued functions of the polar angle $\theta$.
This  fact  and  Eq.~(\ref{III:6a})   tell  us  that  the  eigenvalues  $m$ are
integers for odd winding numbers $n$, and half-integers for even winding numbers
$n$.

Note that, in general,  the  full  system of field  equations (\ref{II:5a}) and
(\ref{II:5b})  cannot   be   described   in   terms   of  the  ansatz  given by
Eqs.~(\ref{II:6}) and (\ref{III:6}).
It can be shown that this  ansatz  is  compatible  with  Eqs.~(\ref{II:5a}) and
(\ref{II:5b}) only under the  condition $\text{Im}\left[ c_{11} c_{12}^{\ast} -
c_{21}c_{22}^{\ast} \right] = 0$.
Obviously,  the  real (modulo  a  common  constant  phase  factor)  radial wave
functions $c_{i a}(r)$ will satisfy this condition.
In particular, the radial wave  functions  of  bound  fermionic states, if any,
can always be chosen to be real.
This is because these wave  functions  satisfy a system  of linear differential
equations with real   $r$-dependent   coefficients and real boundary conditions
$c_{i a}(r) \rightarrow 0$ as $r \rightarrow \infty$.
In contrast, the  radial  wave  functions  of  the  fermionic scattering states
satisfy some  complex  asymptotic  boundary  condition  (i.e., superposition of
incident and  outgoing  waves), and  therefore  do  not  need  to  satisfy  the
reality  condition  $\text{Im}\left[ c_{11} c_{12}^{\ast} - c_{21}c_{22}^{\ast}
\right] = 0$.
When  the  backreaction  terms  are  neglected  in Eq.~(\ref{II:5a}) (as in the
external  field  approximation),  the  ansatz  given  by  Eqs.~(\ref{II:6}) and
(\ref{III:6}) becomes valid without restriction.

We now  discuss  the symmetry properties  of  the  Dirac equation (\ref{III:1})
under discrete transformations.
Unlike the  Dirac  equation  (\ref{A:1})  that  describes  the  states  of free
fermions,  Eq.~(\ref{III:1})   is    not  invariant   under  $C$-transformation
(\ref{A:4}).
Instead,  the  Dirac  equation (\ref{III:1}) transforms into one that describes
fermions in the background field  of  the topological soliton with the opposite
winding number $n$.
The Dirac  equation (\ref{III:1})  is  also  not  invariant  under the modified
$C$-transformation
\begin{equation}
\psi \left( t,\mathbf{x}\right) \rightarrow \psi ^{C^{\prime }}\left(t,
\mathbf{x}\right) =i\gamma ^{1}\mathbb{\otimes }\tau _{2}\psi^{\ast}\left(
t,\mathbf{x}\right),                                             \label{III:6'}
\end{equation}
which  acts on  the  both  spin  and  isospin  indices  of  the  fermion field.
%$\psi \left(t, \mathbf{x} \right)$.
It can easily  be shown that $C^{\prime}$-transformation (\ref{III:6'}) changes
the sign of the Yukawa coupling constant $h$ in Eq.~(\ref{III:2}).
The reason for the noninvariance of Eq.~(\ref{III:1}) under $C'$-transformation
(\ref{III:6'}) is that the fundamental  representation  of the $SU(2)$ group is
pseudoreal.
The   noninvariance   of   Eq.~(\ref{III:1})   under   any   version   of   the
$C$-transformation   leads  to  substantial  differences  between  fermion  and
antifermion scattering in the  background field of the soliton.

Under the inversion $\mathbf{x} \rightarrow -\mathbf{x}$, the isospin component
$\phi_{3}$  of  soliton  solution  (\ref{II:6})  does  not change, whereas both
the $\phi_{1}$ and $\phi_{2}$ components  are  unchanged  for  even $n$ and the
sign changes for odd $n$.
It follows  that  the  Dirac  equation  (\ref{III:1})  is  invariant  under the
$P$-transformation
\begin{equation}
\psi \left( t,\mathbf{x}\right) \rightarrow \psi ^{P}\left( t,\mathbf{x}
\right) =\gamma _{0}\otimes \tau_{3}^{n}\psi \left( t,-\mathbf{x}
\right),                                                        \label{III:6''}
\end{equation}
where $n$ is the winding number of the soliton and we use the fact that $\tau_{
3}^{n}$ is $\mathbb{I}$ for even $n$ and $\tau_{3}$ for odd $n$.
Note that the eigenfunctions $\psi^{\left(m\right)}$  of the grand spin $K_{3}$
are also eigenfunctions of the $P$-transformation:
\begin{equation}
\psi^{\left( m\right) P}=\left(-1\right)^{m\pm \frac{n+1}{2}}\psi
^{\left( m\right)},                                            \label{III:6'''}
\end{equation}
where the upper (lower) sign is used for odd (even) winding numbers $n$.
Hence, unlike the $(3+1)$-dimensional  case,  parity conservation does not lead
to new restrictions on fermion scattering.
This  is  because, unlike  the  $(3 + 1)$-dimensional  case,  the  inversion is
equivalent to a rotation through an angle $\pi$ in $(2 + 1)$ dimensions.
Let us denote by the symbol $\Pi_{2}$ the  operation  of  coordinate reflection
about the  $Ox_{1}$ axis: $\left( x_{1}, x_{2}\right) \rightarrow \left( x_{1},
-x_{2}\right)$.
It can then  easily be shown that the Dirac equation (\ref{III:1}) is invariant
under the combined $\Pi_{2}T$ transformation
\begin{equation}
\psi \left( t,\mathbf{x}\right) \rightarrow \psi^{\Pi_{2}T}\left( t,\mathbf{x}
\right) =\psi ^{\ast }\left( -t,x_{1},-x_{2}\right).          \label{III:6''''}
\end{equation}
Finally, the the Dirac  equation  (\ref{III:1}) is invariant under the combined
transformation
\begin{equation}
\psi \left( t,\mathbf{x}\right) \rightarrow \gamma ^{2}\otimes \tau _{2}\psi
\left( t,x_{1},-x_{2}\right),                                \label{III:6'''''}
\end{equation}
which changes the signs  of  the  orbital, spin,  and  isospin  parts of  grand
spin (\ref{III:4}).
Note  that  the  invariance  of   the   Dirac   equation   (\ref{III:1})  under
transformations  (\ref{III:6''''})  and  (\ref{III:6'''''})  results  from  the
specific property $\boldsymbol{\tau}  \!\boldsymbol{\cdot} \! \boldsymbol{\phi}
\left(x_{1}, -x_{2}\right)  = \boldsymbol{\tau}^{\ast}\!\! \boldsymbol{\cdot}\!
\boldsymbol{\phi} \left(x_{1}, x_{2}\right)$  of  soliton  field  configuration
(\ref{II:6}).

We now turn to the radial dependence  of  the eigenfunctions of the grand spin.
By substituting  Eqs.~(\ref{II:6}), (\ref{II:7}),  and  (\ref{III:6})  into the
Dirac  equation (\ref{III:1}),  we  obtain  the  system  of linear differential
equations for the radial wave functions $c_{i a}(r)$
\begin{equation}
\frac{dc_{ia}}{dr} = \Lambda_{ia;jb}c_{jb},                       \label{III:7}
\end{equation}
where the matrix
\begin{widetext}
\begin{equation}
\Lambda_{ia;jb}=\left(
\begin{array}{cccc}
\dfrac{l_{11}}{r} & 0 & \varepsilon +h-\dfrac{2h}{1+\left( r/r_{0}\right)
{}^{2\left\vert n\right\vert }} & -\dfrac{2h\left( r/r_{0}\right)
{}^{\left\vert n\right\vert }}{1+\left( r/r_{0}\right) {}^{2\left\vert
n\right\vert }} \\
0 & \dfrac{l_{12}}{r} & -\dfrac{2h\left( r/r_{0}\right) {}^{\left\vert
n\right\vert }}{1+\left( r/r_{0}\right) {}^{2\left\vert n\right\vert }} &
\varepsilon -h+\dfrac{2h}{1+\left( r/r_{0}\right) {}^{2\left\vert
n\right\vert }} \\
-\varepsilon +h-\dfrac{2h}{1+\left( r/r_{0}\right) {}^{2\left\vert
n\right\vert }} & -\dfrac{2h\left( r/r_{0}\right) {}^{\left\vert
n\right\vert }}{1+\left( r/r_{0}\right) {}^{2\left\vert n\right\vert }} & -
\dfrac{l_{21}}{r} & 0 \\
-\dfrac{2h\left( r/r_{0}\right) {}^{\left\vert n\right\vert }}{1+\left(
r/r_{0}\right) {}^{2\left\vert n\right\vert }} & -\varepsilon -h+\dfrac{2h}{
1+\left( r/r_{0}\right) {}^{2\left\vert n\right\vert }} & 0 & -
\dfrac{l_{22}}{r}
\end{array}
\right)                                                           \label{III:8}
\end{equation}
\end{widetext}
and $\varepsilon$ is the energy of the fermionic state.

Eqs.~(\ref{III:7})  and  (\ref{III:8})  define  the  system   of   four  linear
differential equations of first order.
In the general case, the  solution  to  this system depends on four independent
parameters.
However, fermionic states should be  described  by  regular solutions to system
(\ref{III:7}).
It can be shown that the  regularity  condition  imposed  at the origin reduces
the number of the independent parameters from four to two.
In the neighborhood  of the origin, a  regular solution to system (\ref{III:7})
has the form
\begin{subequations}                                              \label{III:9}
\begin{eqnarray}
c_{11}\left( r\right)  &=&r^{\left\vert l_{11} \right\vert
}\left( a_{0}+O\left( r^{2}\right) \right),
 \\
c_{12}\left( r\right)  &=&r^{\left\vert l_{12} \right\vert
}\left( b_{0}+O\left( r^{2}\right) \right),
 \\
c_{21}\left( r\right)  &=&r^{\left\vert l_{21}\right\vert
}\left( c_{0}+O\left( r^{2}\right) \right),
 \\
c_{22}\left( r\right)  &=&r^{\left\vert l_{22}\right\vert
}\left( d_{0}+O\left( r^{2}\right) \right),
\end{eqnarray}
\end{subequations}
where only two  of  the  four parameters $a_{0}$, $b_{0}$, $c_{0}$, and $d_{0}$
are independent.
Note that each of the radial  wave functions $c_{ia}(r)$ has  a definite parity
under the substitution $r \rightarrow -r$.
This is because in Eq.~(\ref{III:7}), the differential operator $d/dr$ and each
of the nonzero elements of  the  matrix $\Lambda_{ia; jb}$ also have a definite
parity  under  this substitution.

We present the  linear  relations  between  the  parameters  $a_{0}$,  $b_{0}$,
$c_{0}$, and $d_{0}$ for the  cases of the first odd $(n=1)$ and the first even
$(n=2)$ soliton winding numbers.
By choosing $b_{0}$ and $c_{0}$ as  the  independent  parameters, we obtain for
$a_{0}$ and $d_{0}$:
\begin{eqnarray}
a_{0} &=&-c_{0}\frac{2m}{\varepsilon +h}+\frac{c_{0}}{2}\left( \varepsilon
-h\right) \delta _{m0},                                         \nonumber
\\
d_{0} &=&-b_{0}\frac{\varepsilon -h}{2(m+1)}+c_{0}\frac{2\left\vert
h\right\vert }{r_{0}(\varepsilon +h)}\frac{m}{m+1}               \label{III:10}
\end{eqnarray}
in the case where $n = 1$ and $m \ge 0$, and
\begin{eqnarray}
a_{0} &=&-c_{0}\frac{2\left( m-1/2\right) }{\varepsilon +h}+\frac{c_{0}}{2}
\left( \varepsilon -h\right) \delta _{m\frac{1}{2}},            \nonumber
\\
\;d_{0} &=&-b_{0}\frac{\varepsilon - h}{3 + 2m}+c_{0}\frac{ 2 \left\vert
h\right\vert }{r_{0}^{2}\left( \varepsilon +h\right) }\frac{2m-1}{2m+3}
                                                                 \label{III:11}
\end{eqnarray}
in the case where $n = 2$ and $m \ge 1/2$.
The expressions obtained from  Eqs.~(\ref{III:10})  and  (\ref{III:11})  by the
substitution
\begin{equation}
a_{0}\rightarrow d_{0},\,d_{0}\rightarrow a_{0},\,c_{0}\rightarrow
-b_{0},\,b_{0}\rightarrow -c_{0},\,m\rightarrow - m.             \label{III:12}
\end{equation}
are valid for negative $m$.
Finally, the expressions obtained from  Eqs.~(\ref{III:10}) and  (\ref{III:11})
by the substitution
\begin{equation}
a_{0}\rightarrow b_{0},\, b_{0}\rightarrow a_{0},\, c_{0}\rightarrow
d_{0},\, d_{0}\rightarrow c_{0},\, h \rightarrow -h              \label{III:13}
\end{equation}
can be used in the cases $n = -1$ and $n = -2$, respectively.

It follows from Appendix A that the third component $I_{3}$  of  the isospin is
conserved in the case of free fermions.
The fermion scattering on the soliton can therefore be regarded as a transition
from the free ``in" state,  with  spatial  momentum  $\mathbf{p} = (p, 0)$  and
isospin $I_{3}$, to the  free  ``out" state, with spatial momentum $\mathbf{p}'
= (p\cos(\theta), p\sin(\theta))$ and isospin $I_{3}'$.
A transition without  (with) a change  in  the  isospin can be considered to be
elastic (inelastic).
According to the theory of  scattering \cite{LandauIII, Taylor}, the scattering
state that corresponds to the ``in"  isospin  $I_{3}$  and  the  ``out" isospin
$I_{3}'$  is  described asymptotically by the wave function
\begin{equation}
\Psi _{I_{3}^{\prime },I_{3}} \sim \psi_{\varepsilon,\mathbf{p},I_{3}}\delta_{
I_{3}^{\prime},I_{3}}+\frac{1}{\sqrt{2\varepsilon }}u_{\varepsilon,\mathbf{p}^{
\prime },I_{3}^{\prime }\,}f_{I_{3}^{\prime},I_{3}}\left(p, \theta \right)
\frac{e^{ipr}}{\sqrt{-ir}},                                      \label{III:15}
\end{equation}
where  $\delta_{I_{3}^{\prime},  I_{3}}$   is   the  Kronecker  delta,  $\psi_{
\varepsilon, \mathbf{p}, I_{3}}$  is  wave  function  (\ref{A:6}) of  the ``in"
state with momentum $\mathbf{p}=(p,0)$ and isospin $I_{3}$, $u_{\varepsilon,
\mathbf{p}^{\prime}, I_{3}^{\prime}}$ is the spinor-isospinor amplitude of wave
function (\ref{A:6})  of  the  ``out" state with  momentum $\mathbf{p}^{\prime}
= (p\cos(\theta),  p\sin(\theta))$  and  isospin  $I_{3}^{\prime}$, $f_{I_{3}^{
\prime}, I_{3}}\left(p, \theta\right)$  is the scattering  amplitude,  and  the
 factor $-i$  under  the  square root sign is introduced for convenience.
Using Eqs.~(\ref{A:10}), (\ref{A:14}), (\ref{A:15}), and standard  methods from
the theory of scattering \cite{LandauIII, Taylor}, we  obtain  an expansion  of
the scattering amplitude  $f_{I_{3}^{\prime},I_{3}}\left( p, \theta \right)$ in
terms of the  partial  scattering amplitudes $f_{I_{3}',I_{3}}^{\left(m\right)}
\left( p \right)$
\begin{equation}
f_{I_{3}^{\prime },I_{3}}\left(p, \theta \right) =e^{-iI_{3}^{\prime }\left(
n+1\right) \theta }\sum\limits_{m}f_{I_{3}^{\prime },I_{3}}^{\left( m\right)
}\left( p\right) e^{i m \theta}.                                 \label{III:16}
\end{equation}
In  turn, the partial scattering amplitudes $f_{I_{3}',I_{3}}^{\left(m \right)}
\left( p \right)$  are  expressed  in  terms  of  the  partial  elements of the
$S$-matrix
\begin{subequations}                                             \label{III:17}
\begin{eqnarray}
f_{1/2,1/2}^{\left( m\right) }\left( p\right)  &=&\frac{1}{i\sqrt{2 \pi p}}
\left(S_{1/2,1/2}^{\left( m\right) }\left( p\right) - 1\right), \label{III:17a}
\\
f_{-1/2,1/2}^{\left( m\right) }\left( p\right)  &=&\frac{1}{i\sqrt{2 \pi p}}
S_{-1/2,1/2}^{\left(m\right) }\left( p\right),                  \label{III:17b}
\\
f_{1/2,-1/2}^{\left( m\right) }\left( p\right)  &=&\frac{1}{i\sqrt{2 \pi p}}
S_{1/2,-1/2}^{\left(m\right) }\left( p\right),                  \label{III:17c}
\\
f_{-1/2,-1/2}^{\left( m\right) }\left( p\right)  &=&\frac{1}{i\sqrt{2 \pi p}}
\left(S_{-1/2,-1/2}^{\left(m\right)}\left(p\right)-1\right)\!.  \label{III:17d}
\end{eqnarray}
\end{subequations}
The importance of the partial elements of the $S$-matrix is that the asymptotic
behavior of the radial wave functions $c_{i a}(r)$ can be expressed in terms of
these as $r \rightarrow \infty$:
\begin{widetext}
\begin{equation}
 c_{i a}(r) \sim \frac{\left( -1\right) ^{1/4}}{\sqrt{2\pi pr}}%
\begin{pmatrix}
-i\sqrt{\frac{\varepsilon +h}{2\varepsilon }}\left[ i^{-n}\left( -1\right)
^{m}e^{-ipr}+S_{1/2,1/2}^{\left( m\right) }\left( p\right) e^{ipr}\right]  &
-\sqrt{\frac{\varepsilon -h}{2\varepsilon }}S_{-1/2,1/2}^{\left( m\right)
}\left( p\right) e^{ipr} \\
\sqrt{\frac{\varepsilon -h}{2\varepsilon }}\left[ i^{-n}\left( -1\right)
^{m+1}e^{-ipr}+S_{1/2,1/2}^{\left( m\right) }\left( p\right) e^{ipr}\right]
& -i\sqrt{\frac{\varepsilon +h}{2\varepsilon }}S_{-1/2,1/2}^{\left( m\right)
}\left( p\right) e^{ipr}                                        \label{III:17'}
\end{pmatrix}
\end{equation}
for the ``in" isospin $I_{3} = 1/2$, and
\begin{equation}
c_{i a}(r) \sim \frac{\left( -1\right) ^{1/4}}{\sqrt{2\pi pr}}
\begin{pmatrix}
-i\sqrt{\frac{\varepsilon +h}{2\varepsilon }}S_{1/2,-1/2}^{\left( m\right)
}\left( p\right) e^{ipr} & -\sqrt{\frac{\varepsilon -h}{2\varepsilon }}\left[
i^{n}\left( -1\right) ^{m}e^{-ipr}+S_{-1/2,-1/2}^{\left( m\right) }\left(
p\right) e^{ipr}\right]  \\
\sqrt{\frac{\varepsilon -h}{2\varepsilon }}S_{1/2,-1/2}^{\left( m\right)
}\left( p\right) e^{ipr} & -i\sqrt{\frac{\varepsilon +h}{2\varepsilon }}
\left[ i^{n}\left( -1\right) ^{m+1}e^{-ipr}+S_{-1/2,-1/2}^{\left( m\right)
}\left( p\right) e^{ipr}\right]                                \label{III:17''}
\end{pmatrix}
\end{equation}
for the ``in" isospin $I_{3} = -1/2$.
\end{widetext}

Using standard methods from the theory  of scattering \cite{LandauIII, Taylor},
we obtain an expression for the  differential  scattering  cross-section of the
process $I_{3} \rightarrow I_{3}'$  in  terms  of   the  scattering   amplitude
$f_{I_{3}^{\prime},I_{3}}$
\begin{equation}
d\sigma_{I_{3}^{\prime },I_{3}}/d\theta = \left\vert f_{I_{3}^{\prime
},I_{3}}\left(p, \theta \right) \right\vert^{2}.                 \label{III:18}
\end{equation}
Similarly,  the   partial   cross-sections  of  the  scattering  process $I_{3}
\rightarrow I_{3}'$ are expressed in terms of the partial scattering amplitudes
$f_{I_{3}', I_{3}}^{\left(m \right)}$
\begin{equation}
\sigma _{I_{3}^{\prime },I_{3}}^{\left( m\right) } =
2\pi \left\vert f_{I_{3}^{\prime },I_{3}}^{\left( m\right)}
\left( p \right)\right\vert ^{2}.                               \label{III:18'}
\end{equation}
Note that in  $(2+1)$ dimensions,  the cross-sections $d\sigma_{I_{3}^{\prime},
I_{3}}/d\theta$  and $\sigma _{I_{3}^{\prime }, I_{3}}^{\left( m\right) }$ have
the dimension of length \cite{LandauIII} in the initial dimensional units.

The partial matrix  elements $S_{I_{3}',I_{3}}^{\left( m \right)}$ must satisfy
a  unitarity  condition;  this  follows  from  the unitarity of the $S$-matrix,
$SS^{\dagger } = S^{\dagger }S = \mathbb{I}$,  which  is  a consequence  of the
conservation of probability.
At the same time, the Dirac  equation  results  in  conservation of the fermion
current $j^{\mu} = \bar{\psi}\gamma^{\mu}\psi$, the  time component of which is
the probability density.
Hence, to obtain the  unitarity condition for $S_{I_{3}',I_{3}}^{\left(m\right)
}$ we shall use the conservation  of  the  fermion current: $\partial_{\mu}j^{
\mu} = 0$.
From Eq.~(\ref{III:6}), we  obtain the contravariant components of the partial
fermion  current $j^{\mu \left(m\right)} = \bar{\psi}^{\left( m\right)}\gamma^{
\mu}\psi^{\left( m\right)}$ in polar coordinates:
\begin{subequations}                                             \label{III:19}
\begin{eqnarray}
j^{0\left(m\right)} &=&\left\vert c_{11}\right\vert ^{2}+\left\vert
c_{12}\right\vert^{2}+\left\vert c_{21}\right\vert ^{2}+\left\vert
c_{22}\right\vert^{2},                                          \label{III:19a}
\\
j^{r\left(m\right) } &=&2\,\text{Im}\left[c_{22}c_{12}^{\ast
}-c_{11}c_{21}^{\ast }\right],                                  \label{III:19b}
\\
j^{\theta\left(m\right) } &=&- 2 r^{-1}\text{Re}\left[
c_{11}c_{21}^{\ast }+c_{22}c_{12}^{\ast }\right].               \label{III:19c}
\end{eqnarray}
\end{subequations}
The conservation of $j^{\mu \left(m\right)}$ results  in  the  constancy of the
radial component of the fermion current, i.e. $\partial_{r}j^{r\left(m \right)}
= 0$.
The regularity  of  the  eigenfunctions  $\psi^{\left(m\right)}$  at the origin
leads us  to  the conclusion that  the  radial  component $j^{r\left(m\right)}$
vanishes.
%(\ref{III:17}), and  (\ref{III:18}) together with the expansion (\ref{A:14})
From  Eqs.~(\ref{III:17'}), (\ref{III:17''}),  and  (\ref{III:19b}),  we obtain
the  asymptotic form of the radial component of  the partial fermion current in
terms of the partial matrix elements $S_{I_{3}',I_{3}}^{\left(m\right)}$
\begin{equation}
j^{r \left(m\right)}\sim \frac{v}{2\pi pr}\left(\left\vert S_{1/2,I_{3}}^{
\left( m\right)}\right\vert ^{2}+\left\vert S_{-1/2,I_{3}}^{\left( m\right)}
\right\vert^{2}-1\right),                                        \label{III:20}
\end{equation}
where $I_{3}= \pm 1/2$ is the isospin of the corresponding ``in" fermion state,
and $v = \left(1-h^{2}\varepsilon^{-2}\right)^{1/2}$  is the asymptotic fermion
velocity.
The vanishing of  the  radial  component $j^{r \left(m\right)}$  results in the
diagonal part of  the  unitarity  condition  for  the   partial matrix elements
$S_{I_{3}',I_{3}}^{\left(m\right)}$
\begin{equation}
\sum\limits_{I_{3}^{\prime \prime }}\left[ S_{I_{3}^{\prime \prime
},I_{3}}^{\left( m\right) }\right] ^{\ast }S_{I_{3}^{\prime \prime
},I_{3}^{\prime}}^{\left(m\right)}=\delta_{I_{3},I_{3}^{\prime}}.\label{III:21}
\end{equation}

We now  turn  to  the  symmetry  properties  of  the  partial  matrix  elements
$S_{I_{3}^{\prime },I_{3}}^{\left( m\right)}$.
The basic symmetry property of $S_{I_{3}^{\prime },I_{3}}^{\left( m\right)}$ is
\begin{equation}
S_{-I_{3}^{\prime },-I_{3}}^{\left( -m\right) }=S_{I_{3}^{\prime
},I_{3}}^{\left( m\right)}.                                      \label{III:22}
\end{equation}
This property follows  from  the invariance of the Dirac equation (\ref{III:1})
under  combined   transformation   (\ref{III:6'''''}),   which  in  turn  is  a
consequence of the  specific  form of soliton field configuration (\ref{II:6}).
The sequential action of  the  $\Pi_{2}$  and  $T$  transformations  leaves the
eigenvalue $m$ of grand spin (\ref{III:4}) unchanged  but  permutes the isospin
labels $I_{3}$ and $I_{3}'$.
The invariance   of   the   Dirac   equation   (\ref{III:1})  under  $\Pi_{2}T$
transformation (\ref{III:6''''}) then results in the symmetry relation
\begin{equation}
S_{I_{3}^{\prime },I_{3}}^{\left( m\right) }=S_{I_{3},I_{3}^{\prime
}}^{\left( m\right) }.                                           \label{III:23}
\end{equation}
When $I_{3} = I_{3}'$,  relation  (\ref{III:23})  is  trivial,  but for unequal
(i.e. opposite) $I_{3}$ and $I_{3}'$ it can be rewritten as
\begin{equation}
S_{I_{3}^{\prime },I_{3}}^{\left( m\right) }=S_{-I_{3}^{\prime
},-I_{3}}^{\left( m\right) }\quad \text{if}\quad I_{3}+I_{3}^{\prime} = 0
                                                                 \label{III:24}
\end{equation}
and thus imposes  nontrivial  restrictions  on  $S_{I_{3}^{\prime },I_{3}}^{
\left(m\right)}$.
Finally, from  Eqs.~(\ref{III:22})  and  (\ref{III:24}),  we  obtain a further
symmetry relation for $S_{I_{3}^{\prime },I_{3}}^{\left(m\right)}$
\begin{equation}
S_{I_{3}^{\prime },I_{3}}^{\left( m\right) } = S_{I_{3}^{\prime
},I_{3}}^{\left(-m\right) }\quad \text{if}\quad I_{3}+I_{3}^{\prime } = 0.
                                                                 \label{III:25}
\end{equation}

The partial  elements $S^{\left( m \right)}_{I_{3}',I_{3}}$  of  the $S$-matrix
depend  on the three variables: the fermion momentum $p$,  the  Yukawa coupling
constant $h$, and the  size of the soliton $r_{0}$.
Using    Eq.~(\ref{III:7}),   the    explicit   form   (\ref{III:8})   of   the
$\Lambda$-matrix, and  the  asymptotic  form  (\ref{III:15})  of the scattering
state, it can be shown  that  in  reality, $S^{\left( m\right)}_{I_{3}',I_{3}}$
depend only on two combinations of $p$, $h$, and $r_{0}$:
\begin{equation}
S_{I_{3}^{\prime },I_{3}}^{\left( m\right) } = \xi _{I_{3}^{\prime
},I_{3}}^{\left( m\right) }\left(h^{-1}p, hr_{0}\right),         \label{III:26}
\end{equation}
where $\xi _{I_{3}^{\prime},I_{3}}^{\left( m\right)}$ are some functions of the
two variables.
In particular, the dependence of  $S_{I_{3}^{\prime },I_{3}}^{\left( m\right)}$
on the size of the soliton $r_{0}$ occurs only through the combination $hr_{0}$.

\section{\label{sec:IV} The Born approximation for the scattering amplitudes}

The rather complicated structure of the  matrix  in  Eq.~(\ref{III:8}) makes it
impossible to split system of differential equations (\ref{III:7}) into smaller
subsystems.
Hence, the  solution  to the  system of  four  linear differential equations in
Eq.~(\ref{III:7}) is  reduced to the solution to a linear differential equation
of fourth order.
Due to its rather complex  form,  this  equation cannot be solved analytically.
Consequently, scattering  amplitudes (\ref{III:16})  also cannot be obtained in
an analytical form.
In view of this, it is important  to  investigate the fermion scattering in the
Born  approximation,  which  gives   us  a  chance  to  obtain  an  approximate
analytical expression for the scattering amplitudes $f_{I_{3}^{\prime},I_{3}}$.
%$\left(p,\theta \right)$.

Eq.~(\ref{II:4}) tells us that  the fermion-soliton interaction is described by
the Yukawa term
\begin{equation}
V_{\text{int}} = -h \bar{\psi}\delta \boldsymbol{\phi}
\boldsymbol{\cdot}\boldsymbol{\tau}\psi,                           \label{IV:1}
\end{equation}
where $\delta \boldsymbol{\phi}=\boldsymbol{\phi}-\boldsymbol{\phi}_{\text{v}}$
is the  difference  between  soliton  field  (\ref{II:6})  and the vacuum field
$\boldsymbol{\phi }_{\text{v}} = \left( 0,0,-1\right)$.
The known condition of applicability of the Born approximation \cite{LandauIII}
has the form  $\left\vert V_{\text{int}} \right\vert \ll pa/\left(m_{\psi}a^{2}
\right)$, where $p$ is the momentum of the fermion, $m_{\psi}$ is its mass, and
$a$ is the size  of  the  area  in  which  the  fermion-soliton  interaction is
markedly different from zero.
In our case, this condition can be written as
\begin{equation}
h^{2}r_{0} \ll  p,                                                 \label{IV:2}
\end{equation}
where we have taken into account that in  dimensionless  variables (\ref{II:3})
adopted here, the fermion mass $m_{\psi} = h$.
Note that fulfilling condition (\ref{IV:2}) guarantees  only that the amplitude
of the scattered  outgoing  wave  is  much  smaller  than  the amplitude of the
incident plane wave.
At the same time, for the  Born  approximation  to  be  valid, the second-order
Born amplitude should be much smaller than the first-order one.
We shall see  later  that  for  elastic ($I_{3}' = I_{3}$)  fermion scattering,
this last condition may not be satisfied even if condition (\ref{IV:2}) is met.

Using  standard  methods  from  field  theory  \cite{LandauIV},  we  obtain  an
expression for the first-order Born amplitude  of the scattering process $I_{3}
\rightarrow I_{3}^{\prime}$
\begin{equation}
f_{I_{3}^{\prime },I_{3}} = -\left(8 \pi p\right)^{-1/2}h\bar{u}
_{\varepsilon ,\mathbf{p}^{\prime },I_{3}^{\prime }}\mathbb{I\otimes }\left[
\delta \mathbf{\phi }\left( \mathbf{q}\right) \mathbf{\boldsymbol{\cdot }
\tau }\right] u_{\varepsilon ,\mathbf{p},I_{3}},                   \label{IV:3}
\end{equation}
where
\begin{equation}
\delta \boldsymbol{\phi }\left( \mathbf{q}\right) =
\int \delta \boldsymbol{\phi}
\left( \mathbf{x}\right) e^{-i\mathbf{q x}}d^{2}x                  \label{IV:4}
\end{equation}
and $\mathbf{q}=\mathbf{p}^{\prime } - \mathbf{p}$  is  the  momentum transfer.
The Born amplitudes  can be obtained in an analytical form.
For  winding  numbers   satisfying   the   condition  $\left\vert n \right\vert
\geqslant2$, the amplitudes are expressed in terms of the Meijer $G$ functions.
In the important case when  the  solitons  have  the  winding numbers $n=1$ and
$n=-1$, the Born amplitudes are expressed in terms of modified Bessel functions
of the second kind:
\begin{subequations}                                               \label{IV:5}
\begin{eqnarray}
f_{1/2,1/2} &=&\sqrt{2\pi }hr_{0}^{2}p^{-1/2}( \varepsilon + h
\nonumber \\
&&  -e^{-i\left( \vartheta _{2}-\vartheta _{1} \right) }(\varepsilon
-h) ) \text{K}_{0}\left( qr_{0}\right),                           \label{IV:5a}
  \\
f_{-1/2,1/2} &=&(-1)^{\varkappa}\sqrt{2\pi }hr_{0}^{2}p^{-1/2}\nonumber \\
&& \times e^{i \varkappa \left( \vartheta_{1}+\vartheta _{2}\right) }
q \text{K}_{1}\left( q r_{0}\right),                              \label{IV:5b}
  \\
f_{1/2,-1/2} &=&(-1)^{\varkappa}\sqrt{2\pi }hr_{0}^{2}p^{-1/2}\nonumber \\
&& \times e^{-i \varkappa \left( \vartheta_{1}+\vartheta _{2}\right) }
q \text{K}_{1}\left( q r_{0}\right),                              \label{IV:5c}
  \\
f_{-1/2,-1/2} &=& \sqrt{2\pi }hr_{0}^{2}p^{-1/2}( \varepsilon + h
\nonumber \\
&& -e^{i\left( \vartheta _{2}-\vartheta _{1}\right) }(\varepsilon
-h)) \text{K}_{0}\left( qr_{0}\right),                            \label{IV:5d}
\end{eqnarray}
\end{subequations}
where $\vartheta_{1}$ ($\vartheta_{2}$) is the  polar  angle  that  defines the
direction of motion of the ``in" (``out") fermion, $q=2 p \sin\left( \left\vert
\vartheta_{2} - \vartheta_{1}\right\vert/2\right)$ is the absolute value of the
momentum transfer, and $\varkappa$  is  equal  to $1$ for $n = 1$  and  $0$ for
$n = -1$.
For antifermion scattering, the Born amplitudes take a similar form
\begin{subequations}                                               \label{IV:6}
\begin{eqnarray}
f_{1/2,1/2} &=&-\sqrt{2\pi }hr_{0}^{2}p^{-1/2}( \varepsilon + h
\nonumber \\
&& -e^{i \left( \vartheta _{2}-\vartheta _{1}\right) }(\varepsilon
-h)) \text{K}_{0}\left( qr_{0}\right),                            \label{IV:6a}
  \\
f_{-1/2,1/2} &=&(-1)^{1+\varkappa}\sqrt{2\pi }hr_{0}^{2}p^{-1/2} \nonumber \\
&&\times e^{-i\left( 1-\varkappa \right) \left(\vartheta_{1} + \vartheta
_{2}\right) }q\text{K}_{1}\left( qr_{0}\right),                   \label{IV:6b}
  \\
f_{1/2,-1/2} &=&(-1)^{1+\varkappa}\sqrt{2\pi }hr_{0}^{2}p^{-1/2} \nonumber \\
&&\times e^{i\left( 1-\varkappa \right) \left( \vartheta_{1} + \vartheta
_{2}\right) }q\text{K}_{1}\left( qr_{0}\right),                   \label{IV:6c}
  \\
f_{-1/2,-1/2} &=&-\sqrt{2 \pi} h r_{0}^{2}p^{-1/2}(\varepsilon + h
\nonumber \\
&& -e^{-i \left(\vartheta_{2} - \vartheta_{1}\right)}(\varepsilon
-h)) \text{K}_{0}\left( qr_{0}\right).                            \label{IV:6d}
\end{eqnarray}
\end{subequations}
Note that the  partial elements of the $S$-matrix $S_{I_{3}^{ \prime },I_{3}}^{
\left( m\right)} = i\sqrt{2\pi p} f_{I_{3}^{\prime},I_{3}}^{\left( m \right)} +
\delta _{I_{3}^{\prime },I_{3}}$   that  correspond   to  the  Born  amplitudes
(\ref{IV:5}) and (\ref{IV:6}) can be written in form (\ref{III:26}).

From Eqs.~(\ref{IV:5}) and (\ref{IV:6})  it  follows  that  the  amplitudes are
Hermitian with respect to the permutation of the isospins  and  momenta  of the
fermionic states
\begin{equation}
f_{I_{3}^{\prime },I_{3}}\left( \vartheta ^{\prime },\vartheta \right)
=f_{I_{3},I_{3}^{\prime }}^{\ast }\left( \vartheta ,\vartheta ^{\prime
}\right)                                                           \label{IV:7}
\end{equation}
as required for the Born approximation \cite{LandauIII, Taylor}.
Next, the  invariance  of  Eq.~(\ref{III:1})  under   $\Pi_{2}T$ transformation
(\ref{III:6''''})  leads   to  a  symmetry  relation  for  the  Born amplitudes
\begin{equation}
f_{I_{3}^{\prime },I_{3}}\left(\vartheta', \vartheta \right)
= f_{I_{3},I_{3}^{\prime}}\left(\pi - \vartheta, \pi - \vartheta'\right).
                                                                  \label{IV:7'}
\end{equation}
Another symmetry  relation
\begin{equation}
f_{I_{3}^{\prime },I_{3}}\left( \vartheta', \vartheta \right)
= f_{-I_{3}^{\prime },-I_{3}}\left(-\vartheta', -\vartheta \right)
                                                                 \label{IV:7''}
\end{equation}
follows from the  invariance of Eq.~(\ref{III:1}) under combined transformation
(\ref{III:6'''''}).
Note that in  Eqs.~(\ref{IV:7'}) and (\ref{IV:7''}), the minus signs before the
angular  variables  are  due  to  the  fact  that  the $\Pi_{2}$ transformation
changes the signs  of  the  $y$ components  of  the  momenta  of  the fermions.
Finally, Eqs.~(\ref{IV:5}) and (\ref{IV:6})  tell  us  that  scattering  of the
fermion in the background field of the  soliton  with  winding  number $n = \pm
1$ is equivalent to  scattering  of  the  antifermion  in  the background field
of the soliton with the opposite winding number.
We use  the  superscript $[n,+]$  ($[n,-]$)  to  indicate the scattering of the
fermion  (antifermion) in  the  background  field  of  the soliton with a given
winding number $n$.
It  then  follows  from   Eqs.~(\ref{IV:5})  and  (\ref{IV:6})  that  the  Born
amplitudes satisfy the relations:
\begin{subequations}                                               \label{IV:8}
\begin{eqnarray}
f_{I_{3}^{\prime },I_{3}}^{\left[ \pm 1,+\right] }\left( \vartheta
_{2},\vartheta _{1}\right)  &=&\left( -1\right) ^{I_{3}^{\prime
}+I_{3}}f_{-I_{3}^{\prime },-I_{3}}^{\left[ \mp 1,-\right] }\left( \vartheta
_{2},\vartheta_{1}\right),                                        \label{IV:8a}
 \\
f_{I_{3}^{\prime },I_{3}}^{\left[ \pm 1,+\right] }\left( \vartheta
_{2},\vartheta _{1}\right)  &=&f_{I_{3}^{\prime },I_{3}}^{\left[ \pm 1,+
\right] }\left( \vartheta _{2}+\pi ,\vartheta _{1}+\pi \right),   \label{IV:8b}
 \\
f_{I_{3}^{\prime },I_{3}}^{\left[ \pm 1,-\right] }\left( \vartheta
_{2},\vartheta _{1}\right)  &=&f_{I_{3}^{\prime },I_{3}}^{\left[ \pm 1,-
\right] }\left( \vartheta _{2}+\pi ,\vartheta _{1}+\pi \right).   \label{IV:8c}
\end{eqnarray}
\end{subequations}
Eq.~(\ref{IV:8a})  is  a  consequence  of  the fact that the winding number $n$
in the Dirac equation (\ref{III:1}) changes the  sign  under $C$-transformation
(\ref{A:4}).
Eqs.~(\ref{IV:8b})  and  (\ref{IV:8c}) follow  from the invariance of the Dirac
equation (\ref{III:1}) under $P$-transformation (\ref{III:6''}).

We can now ascertain the behavior of the Born amplitudes (\ref{IV:5}) for large
and small values of the  momentum transfer $q$.
Using the known asymptotic forms of the modified Bessel functions $\text{K}_{0}
\left(q r_{0} \right)$   and   $\text{K}_{1}\left( q r_{0}\right)$,  we  obtain
asymptotic forms  of  the  Born amplitudes (\ref{IV:5}) at large values of $q$:
\begin{subequations}                                               \label{IV:9}
\begin{eqnarray}
f_{\pm 1/2,\pm 1/2} &\sim &2^{-1/2}\pi hr_{0}^{3/2}e^{-qr_{0}}\left(
1-e^{\mp i\left( \vartheta _{2}-\vartheta _{1}\right) }\right) \nonumber  \\
&&\times \sin \left( \left\vert \vartheta _{2}-\vartheta _{1}\right\vert
/2\right) ^{-1/2}, \\
f_{\mp 1/2,\pm 1/2} &\sim &(-1)^{\varkappa }\sqrt{2}\pi
hr_{0}^{3/2}e^{-qr_{0}}e^{\pm i\varkappa \left( \vartheta _{2}+\vartheta
_{1}\right) } \nonumber \\
&&\times \sin \left( \left\vert \vartheta _{2}-\vartheta _{1}\right\vert
/2\right) ^{1/2},
\end{eqnarray}
\end{subequations}
where the angles $\vartheta_{1}$ and $\vartheta_{2}$ are fixed and $\vartheta_{
1} \ne \vartheta_{2}$.
We see that the Born amplitudes decrease exponentially with an increase in both
the momentum transfer $q$ and the effective size of the soliton $r_{0}$.

Next we consider the case of low momentum transfer  $q$  and high fixed fermion
momentum $p$.
This situation  involves small scattering angles $\Delta \vartheta \equiv \left
\vert \vartheta_{2} - \vartheta_{1}\right\vert = 2 \arcsin \left[ q/ \left( 2 p
\right) \right] \approx q/p$.
In this case, the Born amplitudes take the form
\begin{subequations}                                              \label{IV:10}
\begin{align}
f_{\pm 1/2,\pm 1/2} &\sim -2\sqrt{2\pi}h^{2}r_{0}^{2}p^{-1/2}\left(\ln
\left( qr_{0}/2\right)+\gamma \right),                           \label{IV:10a}
\\
f_{\mp 1/2,\pm 1/2} &\sim (-1)^{\varkappa }\sqrt{2\pi}
hr_{0}p^{-3/2}\left( p\pm i\varkappa q\right),                   \label{IV:10b}
\end{align}
\end{subequations}
where $\gamma$ is the Euler-Mascheroni constant.
It follows from  Eq.~(\ref{IV:10a}) that in the elastic channel (with no change
in isospin), the Born amplitudes diverge logarithmically  as $q \rightarrow 0$.
Conversely, the  inelastic  Born  amplitudes  tend  to  constant values in this
limit.
More  importantly, the elastic  Born  amplitudes  $\propto h^{2}$,  whereas the
inelastic   ones  $\propto  h$,  as   expected   for   the   usual  first  Born
approximation in which amplitudes are proportional to a coupling constant.
The reason for this  behavior  of  the  elastic  Born  amplitudes  lies  in the
spin-isospin structure of plane-wave  fermionic states (\ref{A:6}) entering the
Born amplitudes (\ref{IV:3}), and in the mechanism of generation of the fermion
mass in model (\ref{II:1}).

In model (\ref{II:1}), fermions gain mass due  to   spontaneous breaking of the
$SU(2)$ global symmetry.
In our dimensionless notation (\ref{II:3}), the fermion mass $m_{\psi}$ is equal
to the Yukawa coupling constant $h$.
Next, relativistic invariance results in the factors $(\varepsilon \pm m_{\psi}
)^{1/2} = (\varepsilon \pm h)^{1/2}$  in  fermion  wave  functions (\ref{A:6}).
In turn, these factors and the spin-isospin structure of Eq.~(\ref{IV:3}) result
in the characteristic factors $\left[\varepsilon+h- e^{\mp i\left(\vartheta_{2}
- \vartheta_{1}\right)}(\varepsilon - h)\right]$  in  elastic  Born  amplitudes
(\ref{IV:5a}),  (\ref{IV:5d}),  (\ref{IV:6a}),  and (\ref{IV:6d}), whereas such
factors are absent from inelastic Born amplitudes (\ref{IV:5b}), (\ref{IV:5c}),
(\ref{IV:6b}),  and  (\ref{IV:6c}).
For  small scattering  angles $\Delta\vartheta = \vartheta_{2} - \vartheta_{1}$
and large  fermion  momenta $p$,  these  factors  take  the  form $ 2 h \pm i p
\Delta \vartheta$.
We can see that for scattering angles $\Delta \vartheta < 2 h p^{-1}$, the term
$2h$ becomes predominant.
Hence, the terms that $\propto h^{2}$  also  become predominant in elastic Born
amplitudes  (\ref{IV:5a}),   (\ref{IV:5d}),  (\ref{IV:6a}),  and  (\ref{IV:6d})
that were obtained  within  the  framework  of  the  first  Born approximation.
It  follows   that   elastic   Born  amplitudes  (\ref{IV:5a}),  (\ref{IV:5d}),
(\ref{IV:6a}),  and  (\ref{IV:6d})  become  inapplicable  when  the  scattering
angle $\Delta \vartheta \lesssim 2 h p^{-1}$.
This is because the contribution  of  the  second  Born  approximation  to  the
scattering amplitudes is $\propto h^{2}$,  and  is therefore  of the same order
of magnitude  as the contribution  of  the  $h^{2}$ terms  of  the  first  Born
approximation.
Note that for large fermion momenta $p$,  the  domain  of $\Delta \vartheta$ in
which the scattering amplitudes  are  markedly  different  from  zero is of the
order of $(p r_{0})^{-1}$.
It follows that  under  the   condition $2 h r_{0} \sim 1$,  the  area  $\Delta
\vartheta  \lesssim 2 h p^{-1}$  in  which  the  first  Born  approximation  is
incorrect may cover a substantial part  of the area  $\Delta \vartheta \lesssim
(p r_{0})^{-1}$ in which the  scattering amplitudes are markedly different from
zero, and may even overlap it completely.
It was found that in the case of solitons with higher winding  numbers $n$, the
characteristic  factor  $\left[\varepsilon  + h - e^{\mp i \left(\vartheta_{2}-
\vartheta_{1}\right)}(\varepsilon - h)\right]$   also  arises  in  the  elastic
first-order Born  amplitudes,  meaning  that these are  inapplicable  when  the
scattering angle $\Delta \vartheta \lesssim 2 h p^{-1}$.

Eqs.~(\ref{III:16}), (\ref{IV:5}), and (\ref{IV:6}) make  it possible to obtain
analytical expressions  for  the  partial  amplitudes  $f_{I_{3}^{\prime }, I_{
3}}^{\left( m \right) }\left( p \right)$  in terms  of  the Meijer G functions.
As $p \rightarrow \infty$,  these  expressions  tend  to  the asymptotic forms:
\begin{subequations}                                             \label{IV:10'}
\begin{eqnarray}
f_{\pm 1/2,\pm 1/2}^{\left( m\right) } &\sim &\sqrt{2 \pi} h r_{0} p^{-1/2}
                                                                \label{IV:10'a}
\\
&&\times \left[ hp^{-1}+2^{-2}p^{-2}r_{0}^{-2}\left( 1-2\left( \varkappa \mp
m\right) \right) \right],
\nonumber \\
f_{\pm 1/2,\mp 1/2}^{\left( m\right) } &\sim &\left( -1\right)^{\varkappa
}\sqrt{\pi /2}hp^{-1/2}                                         \label{IV:10'b}
\\
&&\times \left[ p^{-1}+2^{-3}\left( 1-4m^{2}\right) p^{-3}r_{0}^{-2}\right]
\nonumber
\end{eqnarray}
\end{subequations}
for fermion scattering, and
\begin{subequations}                                            \label{IV:10''}
\begin{eqnarray}
f_{\pm 1/2,\pm 1/2}^{\left( m\right) } &\sim &-\sqrt{2\pi }hr_{0}p^{-1/2} \\
&&\times \left[ hp^{-1}+2^{-2}p^{-2}r_{0}^{-2}\left( 1-2\left( \varkappa \pm
m\right) \right) \right],
\nonumber \\
f_{\pm 1/2,\mp 1/2}^{\left( m\right) } &\sim &\left( -1\right) ^{1+\varkappa}
\sqrt{\pi /2}hp^{-1/2} \\
&&\times \left[ p^{-1}+2^{-3}\left( 1-4 m^{2}\right) p^{-3}r_{0}^{-2}\right]
\nonumber
\end{eqnarray}
\end{subequations}
for antifermion scattering.
It follows  from  Eq.~(\ref{III:17})  that  the  partial amplitudes $f_{I_{3}^{
\prime},I_{3}}^{\left(m\right)}$  must  satisfy  the same symmetry relations as
the  partial  matrix  elements  $S_{I_{3}^{\prime },I_{3}}^{\left( m \right)}$,
and  we  can  see  that  the   Born   partial   amplitudes  (\ref{IV:10'})  and
(\ref{IV:10''})  satisfy   symmetry   relations (\ref{III:22}), (\ref{III:23}),
(\ref{III:24}), and (\ref{III:25}).
Eqs.~(\ref{IV:10'})  and  (\ref{IV:10''})  tell  us  that the leading  terms of
the asymptotic  forms  of  $f_{I_{3}^{\prime},I_{3}}^{\left(m\right)}$  do  not
depend on $m$.
Furthermore, the  leading  terms  of  the  elastic  ($I_{3} = I_{3}'$)  partial
amplitudes are $\propto h^{2}$, whereas those of the inelastic ($I_{3} \ne I_{3
}'$) ones are $\propto h$.
Finally,  we  recall   that  the  Born  partial  amplitudes  and  corresponding
partial elements of the $S$-matrix  do   not   satisfy  the unitarity condition
\cite{LandauIII, Taylor}. %(\ref{III:21}).

We can  also  determine  the  asymptotic  behavior  of  the  partial amplitudes
$f_{I_{3}^{\prime},I_{3}}^{\left( m \right)}\left( p \right)$  as $\left\vert m
\right\vert\rightarrow\infty$.
Using  known   methods   \cite{Olver}  for   the   calculation  of  the Fourier
coefficients in the limit of large $m$, we obtain the corresponding  asymptotic
forms of the Born partial amplitudes:
\begin{subequations}                                           \label{IV:10'''}
\begin{eqnarray}
f_{\pm 1/2,\pm 1/2}^{\left( m\right) } &\sim &\sqrt{\pi /2}hr_{0}^{2}p^{-1/2}
\left[ \left( \varepsilon +h\right) \left\vert m\mp \varkappa \right\vert
^{-1}\right. \nonumber \\
&&\left. -\left( \varepsilon -h\right) \left\vert m\mp  \varkappa
\pm 1 \right\vert ^{-1}\right], \\
f_{\pm 1/2,\mp 1/2}^{\left( m\right) } &\sim &\left( -1\right) ^{\varkappa
}\sqrt{\pi /2}hr_{0}^{3}p^{3/2}\left\vert m\right\vert ^{-3}
\end{eqnarray}
\end{subequations}
for fermion scattering, and
\begin{subequations}                                          \label{IV:10''''}
\begin{eqnarray}
f_{\pm 1/2,\pm 1/2}^{\left( m\right) } &\sim &-\sqrt{\pi /2}hr_{0}^{2}p^{-1/2}%
\left[ \left( \varepsilon +h\right) \left\vert m\pm \varkappa \mp
1\right\vert ^{-1}\right.  \nonumber \\
&&\left. -\left( \varepsilon -h\right) \left\vert m\pm \varkappa \right\vert
^{-1}\right] , \\
f_{\pm 1/2,\mp 1/2}^{\left( m\right) } &\sim &\left( -1\right) ^{1+\varkappa
}\sqrt{\pi /2}hr_{0}^{3}p^{3/2}\left\vert m\right\vert ^{-3}
\end{eqnarray}
\end{subequations}
for antifermion scattering.

From the  analytical  expressions  for  the Born   amplitudes  (\ref{IV:5}) and
Eq.~(\ref{III:18}),  we  obtain  differential scattering cross-sections for the
processes $I_{3} \rightarrow I_{3}'$ in the Born approximation
\begin{subequations}                                              \label{IV:11}
\begin{eqnarray}
d\sigma _{\pm 1/2,\pm 1/2}/d\vartheta  &=&8\pi p^{-1}h^{2}r_{0}^{4}\text{K}
_{0}\left( 2pr_{0}\sin \left( \vartheta /2\right) \right) ^{2} \nonumber \\
&&\times \left( h^{2}+p^{2}\sin \left( \vartheta /2\right)^{2}
\right),                                                         \label{IV:11a}
\end{eqnarray}
\begin{eqnarray}
d\sigma _{\mp 1/2,\pm 1/2}/d\vartheta  &=&8\pi ph^{2}r_{0}^{4}\text{K}
_{1}\left( 2pr_{0}\sin \left( \vartheta /2\right) \right) ^{2} \nonumber \\
&&\times \sin \left( \vartheta /2\right)^{2}.                    \label{IV:11b}
\end{eqnarray}
\end{subequations}
Since the corresponding  Born  amplitudes in Eqs.~(\ref{IV:5}) and (\ref{IV:6})
differ  only  by  a  phase  factor, the  Born  differential  cross-sections for
antifermion scattering are the same as those for fermion scattering.
Note  that  in  accordance  with  Eq.~(\ref{IV:10}), the  elastic  differential
cross-sections diverge logarithmically according to the leading term $8\pi h^{4
}r_{0}^{4}p^{-1}\left( \gamma +\ln \left( pr_{0}\vartheta/2\right)\right)^{2}$,
whereas the inelastic  ones  tend  to  a constant value of $2\pi h^{2}r_{0}^{2}
p^{-1}$  as  $\vartheta \rightarrow 0$ for large but fixed $p$.
The total cross-sections of  the  processes $I_{3} \rightarrow I_{3}'$ can also
be obtained in analytical form
\begin{subequations}                                              \label{IV:12}
\begin{eqnarray}
\sigma _{\pm 1/2,\pm 1/2} &=&8\pi ^{2}p^{-1}h^{2}r_{0}^{4} \nonumber \\
&&\times \left\{ p^{2}G_{2,4}^{3,1}\left( 4p^{2}r_{0}^{2}\left\vert
\begin{array}{c}
-\frac{1}{2},\frac{1}{2} \\
0,0,0,-1
\end{array}
\right. \right) \right.  \nonumber \\
&&\left. +h^{2}G_{2,4}^{3,1}\left( 4p^{2}r_{0}^{2}\left\vert
\begin{array}{c}
\frac{1}{2},\frac{1}{2} \\
0,0,0,0
\end{array}
\right. \right) \right\},                                        \label{IV:12a}
\end{eqnarray}
\begin{equation}
\sigma _{\mp 1/2,\pm 1/2}=8\pi ^{2}ph^{2}r_{0}^{4}G_{2,4}^{3,1}\left(
4p^{2}r_{0}^{2}\left\vert
\begin{array}{c}
-\frac{1}{2},\frac{1}{2} \\
-1,0,1,-1
\end{array}
\right. \right),                                                 \label{IV:12b}
\end{equation}
\end{subequations}
where $G^{m,n}_{p,q}$  are  the  Meijer  $G$  functions, defined  according  to
Ref.~\cite{PBM}.
Finally, using known asymptotic  expansions  for  the Meijer  $G$  functions, we
obtain the  asymptotics  of  the  total  cross-sections  (\ref{IV:12}) for large
values of the fermion momenta $p$
\begin{subequations}                                              \label{IV:13}
\begin{eqnarray}
\sigma _{\pm 1/2,\pm 1/2} &=&\frac{1}{8}\pi ^{3}h^{2}r_{0}       \label{IV:13a}
\\
&&\times \left( \left( 1+32h^{2}r_{0}^{2}\right) p^{-2}+O\left(
p^{-4}\right) \right), \nonumber \\
\sigma _{\mp 1/2,\pm 1/2} &=&\frac{3}{8}\pi ^{3}h^{2}r_{0}\left(
p^{-2}+O\left( p^{-4}\right) \right).                            \label{IV:13b}
\end{eqnarray}
\end{subequations}
Note that  Eqs.~(\ref{IV:11a}), (\ref{IV:12a}),  and  (\ref{IV:13a}), which are
related  to  the  elastic  channel, include  terms  with an  additional  factor
$h^{2}$ in comparison with the usual first Born approximation.

It  follows  from   Eqs.~(\ref{IV:12})  and  (\ref{IV:13})  that  in  the  Born
approximation, the total cross-section of the fermion scattering, which is equal
to the sum of the elastic (\ref{IV:12a}) and inelastic (\ref{IV:12b}) parts, is
finite.
Next,   it   follows   from    Eq.~(\ref{IV:5})  that   for   small  scattering
angles, the imaginary parts of the elastic Born amplitudes $f_{\pm 1/2,\pm 1/2}
$ have the form  $\text{Im}\left[f_{\pm 1/2,\pm 1/2}\right] \sim \mp\sqrt{2\pi}
h r_{0}^{2}  p^{-1/2} \left( \varepsilon - h \right) \left( \gamma + \ln \left[
p r_{0}\vartheta /2 \right] \right) \vartheta$, and consequently vanish at zero
scattering angle.
At the same time, in our case, the optical theorem of the scattering theory can
be written as
\begin{equation}
\text{Im}\left[ f_{I_{3},I_{3}}\left( p,\vartheta =0\right) \right] =
\frac{1}{2}\sqrt{\frac{p}{2\pi }}\left( \sigma _{1/2,I_{3}}\left( p\right)
+ \sigma_{-1/2,I_{3}}\left( p\right) \right).                     \label{IV:14}
\end{equation}
We see  that  the  optical  theorem  is  not  valid  in the Born approximation.
This is because the optical theorem is a consequence  of  the  unitarity of the
$S$-matrix, which is violated in the Born approximation.

\section{\label{sec:V} Numerical results}

The radial wave  functions  of fermionic scattering states  are  solutions to
system (\ref{III:7}) that satisfy regularity condition (\ref{III:9}).
As $r \rightarrow \infty$, the radial wave functions $c_{i a}(r)$ tend to their
asymptotic forms (\ref{III:17'})  and  (\ref{III:17''}), which are expressed in
terms of the  partial elements of the $S$-matrix (\ref{III:17}).
The components  of  Eq.~(\ref{III:17'}) that  correspond  to the ``out" isospin
$I_{3}'=-1/2$ and those of Eq.~(\ref{III:17''}) that  corresponds to the ``out"
isospin $I_{3}'=1/2$  contain  only  outgoing  waves, and  this should be clear
from the physical background.

Our goal is to find the partial matrix elements $S^{(m)}_{I_{3}',I_{3}}(p)$ for
a range of fermion momenta $p$, as this will give the most complete description
of the fermion scattering.
To  do  this, we  numerically  solve  the  system  in  Eq.~(\ref{III:7})  under
regularity  condition (\ref{III:9}) for  a set of $p$ within some finite range.
%We shall consider only the important case of the ``elementary" soliton with the
%unit winding number.
The origin $r =0$  in the neighborhood of which regular expansion (\ref{III:9})
is  valid  is the regular singular point of the system in Eq.~(\ref{III:7}).
Hence, the  initial  value  problem  (IVP)  for  system (\ref{III:7}) cannot be
posed at $r=0$.
To work around this problem, we shift the point at which the  IVP is posed to a
short distance from the origin.
The initial values of  the  radial  wave  functions $c_{i a}(r)$ at the shifted
initial  point  $r_{0}$   are   calculated   using   regular   power  expansion
(\ref{III:9}).

A regular  solution  to  the  system  in  Eq.~(\ref{III:7})  depends on the two
complex parameters $b_{0}$ and $c_{0}$.
The linearity of the system makes it possible to  set  one  of these parameters
(for instance $b_{0}$) to one.
In this way, we  stay  with  the  two  real  parameters $\text{Re}[c_{0}]$ and
$\text{Im}[c_{0}]$.
Consider fermion scattering  that corresponds to the ``in" fermionic state with
isospin $I_{3} = 1/2$.
To obtain the  radial  wave  functions  $c_{i a}(r)$  that  correspond  to this
scattering state, we must satisfy the condition  that the radial wave functions
$c_{i 2}(r)$ contain only the outgoing  wave  as  it is in Eq.~(\ref{III:17'}).
It follows from  Eq.~(\ref{III:17'}) that for large $r$, the real and imaginary
parts of the outgoing radial wave functions  $c_{i 2}(r)$  should  satisfy  the
asymptotic  condition: $\text{Im}[ \sqrt{pr}c_{i2}(p r + \delta_{2} + \pi/2)] =
\text{Re}[\sqrt{p r} c_{i2}(p r + \delta_{2})]$,  where  $\delta_{2}$  is  some
phase shift.
To satisfy this condition, we can use the two parameters $\text{Re}[c_{0}]$ and
$\text{Im}[c_{0}]$.
By varying  $\text{Re}[c_{0}]$   and  keeping  $\text{Im}[c_{0}]$ fixed, we can
achieve coincidence between the zeros of $\text{Im}[\sqrt{pr}c_{i2}(pr+\delta_{
2}+\pi/2)]$ and $\text{Re}[\sqrt{pr} c_{i2}(p r +\delta_{2})]$.
Next, by  varying   $\text{Im}[c_{0}]$  and keeping $\text{Re}[c_{0}]$ fixed,
we can achieve equality of $\text{Im}[\sqrt{pr}c_{i2}(pr +\delta_{2} + \pi/2)]$
and $\text{Re}[\sqrt{pr} c_{i2}(pr + \delta_{2})]$.
The positions of the zeros of $\text{Im}[\sqrt{pr}c_{i2}(pr+\delta_{2}+\pi/2)]$
and $\text{Re}[\sqrt{p r}c_{i2}(p r +\delta_{2})]$  do  not  change  under this
variation, because linear system (\ref{III:7}) contains  only  real coefficient
functions and thus  the  imaginary  parts of $c_{i a}(r)$ are $\propto \text{Im
}[c_{0}]$, since $i \text{Im}[c_{0}]$ is  the  only  imaginary parameter of the
problem.
In turn, linear scaling  cannot change the location of the zeros of  $\text{Im}
[c_{i a}(r)]$.
When the asymptotic  condition  is  satisfied,  we  can assume  that the radial
wave functions  $c_{i a}(r)$  correspond  to  the ``in"  fermionic  state  with
isospin $I_{3} = 1/2$.
Note  that  the  asymptotic condition  for   the  ``in"  fermionic  state  with
isospin  $I_{3} = -1/2$  has  a similar form: $\text{Im}[\sqrt{pr}c_{i 1}(p r +
\delta_{1} + \pi/2)] = \text{Re}[\sqrt{p r} c_{i 1}(p r + \delta_{1})]$.
The value of $r$ at which  the  asymptotic  condition  is  satisfied  should be
chosen based on  the  accuracy  of  asymptotic  form (\ref{III:17'}) and on the
low influence of the background field of the soliton.

To  solve  the  IVP  numerically,  we  use  the  IVP  solver  provided  by  the
{\sc{Maple}} package \cite{maple}.
This  solver  finds  a  numerical  solution  to  the  IVP  using  the  Fehlberg
fourth-fifth order Runge-Kutta method with degree four interpolant.
Having obtained the solution to the IVP, we can  determine  the coefficients of
the outgoing ($\propto \exp(i p r)$) and ingoing ($\propto \exp(-i p r)$) parts
of the radial wave functions $c_{i a}(r)$.
To do this, we use the orthonormality  property  of  the  exponential functions
$\int\nolimits_{0}^{2\pi p^{-1}}e^{\kappa_{1} i p r} e^{\kappa_{2} i p r}dr = 2
\pi p^{-1} \delta _{0,\kappa_{1} + \kappa _{2}}$, where $\kappa_{1,2} = \pm 1$.
Taking into account the phase and normalization factors in Eq.~(\ref{III:17'}),
we obtain the values  of  the  partial  matrix  elements $S_{I_{3}^{\prime},I_{
3}}^{\left( m\right)}$ for a given $p$.
To control  the  correctness  of  the  numerical  solution  to  the IVP, we use
unitarity condition (\ref{III:21}) and the equality of the pairs of $S_{I_{3}^{
\prime },I_{3}}^{\left( m\right)}$ obtained from  the up and down components of
the columns in Eq.~(\ref{III:17'}).% and (\ref{V:2}).

We now turn to a discussion of the numerical results.
We consider only  fermion  scattering  on  the elementary ($n = 1$) topological
soliton of the nonlinear $O(3)$ $\sigma$-model.
The scattering corresponds to the ``in" fermionic state  with  isospin $I_{3} =
1/2$; from Eqs. (\ref{III:22}), it follows that  scattering  with ``in" isospin
$I_{3}=-1/2$ is in essence equivalent to the case considered here.
The size parameter $r_{0}$ and the Yukawa  coupling  constant $h$  are the only
parameters of the IVP.
In order for the external  field  approximation  to  be valid, the  mass of the
fermion, which is equal to $h$  in  dimensionless units (\ref{II:3}), should be
much lower than the  mass of the soliton $4 \pi$.
Hence,  we  set  the  parameters  $h$  and  $r_{0}$  equal  to  $0.1$  and $1$,
respectively.
According to Eq.~(\ref{III:26}), the numerical solution  corresponding to these
values $h$ and $r_{0}$ will also give us information about the partial elements
of the $S$-matrix for which the parameters satisfy the condition $h r_{0}= 0.1$
The presence  of  the  inelastic  scattering  channel $I_{3}\rightarrow I_{3}'$
($I_{3} \ne I_{3}'$) results  in  $S_{1/2, 1/2}^{\left( m \right)}$ leaving the
unitary circle, and thus the description of the scattering in  terms of a phase
shift loses its advantage.
In view of this, we shall describe the  partial  elements  of the $S$-matrix in
terms of their real and imaginary parts.

\begin{figure}[tbp]
\includegraphics[width=0.5\textwidth]{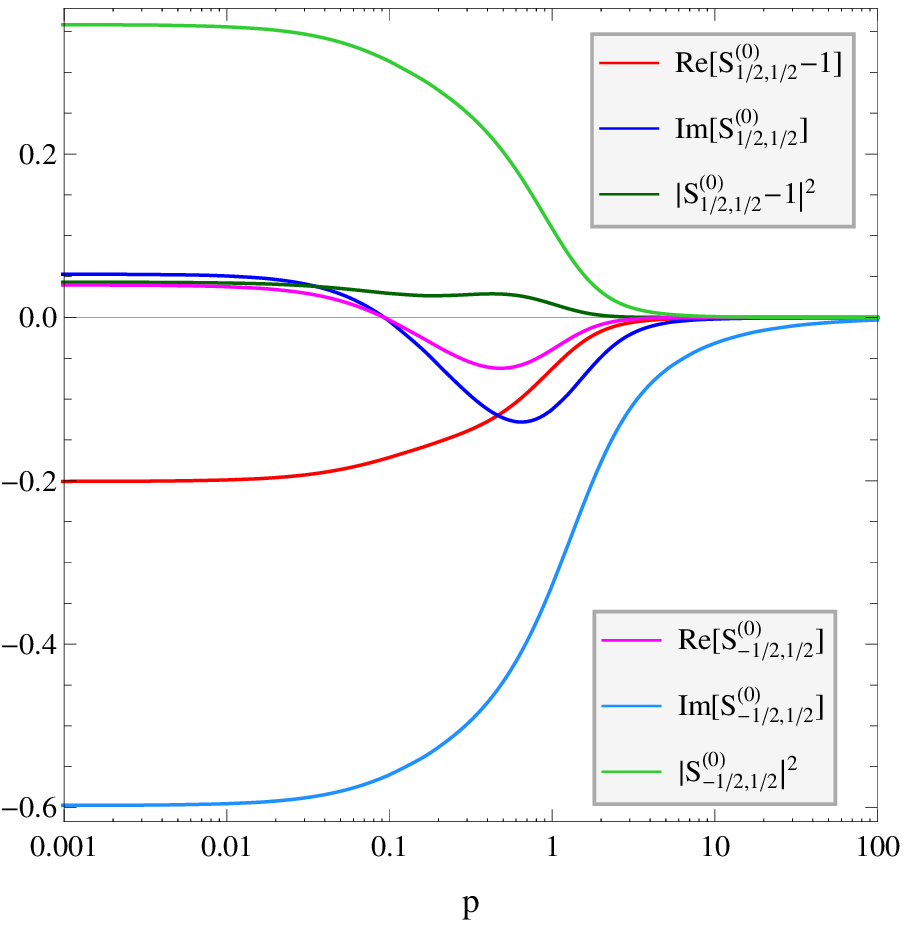}
\caption{\label{fig1}      Dependence of $\text{Re}[S_{1/2,1/2}^{\left(0\right)
}-1]$, $\text{Im}[S_{1/2,1/2}^{\left( 0\right)}]$, $\vert S_{1/2, 1/2}^{\left(0
\right)}\vert^{2}$, $\text{Re}[S_{-1/2,1/2}^{\left( 0\right)}]$, $\text{Im}[S_{
-1/2,1/2}^{\left(0\right)}]$, and $\vert S_{-1/2,1/2}^{\left( 0 \right)}\vert^{
2}$ on the fermion momentum $p$.}
\end{figure}

It was  found  that  the   $p$-dependences   of  the   partial  matrix elements
$S_{I_{3}^{\prime},I_{3}}^{\left(0 \right)}$, $S_{I_{3}^{\prime},I_{3}}^{\left(
\pm 1 \right)}$, and $S_{I_{3}^{\prime},I_{3}}^{\left( \left\vert m \right\vert
\ge 2 \right)}$ are substantially  different,  and  we therefore  consider them
separately here.
For better visualization  of  the  $p$-dependences,  we show them on log-linear
plots.
Figure~\ref{fig1}  shows  the  $p$-dependences  related  to  the partial matrix
elements $S_{\pm 1/2,1/2}^{\left(0\right)}$.
We can see that all of the curves in  Fig.~\ref{fig1}  are regular functions of
the fermion momentum $p$.
In  particular,  they   all   tend  to  nonzero  limits  as  $p \rightarrow 0$,
and $\vert S_{1/2,1/2}^{\left( 0 \right)} - 1 \vert^{2}$, $\vert S_{-1/2,1/2}^{
\left( 0 \right) }\vert^{2}$, $\vert \text{Re}[S_{1/2,1/2}^{\left(0\right) }-1]
\vert$,  and  $\vert \text{Im}[S_{-1/2, 1/2}^{\left( 0 \right)}]  \vert$  reach
their maximal values at $p = 0$.

\begin{figure}[tbp]
\includegraphics[width=0.5\textwidth]{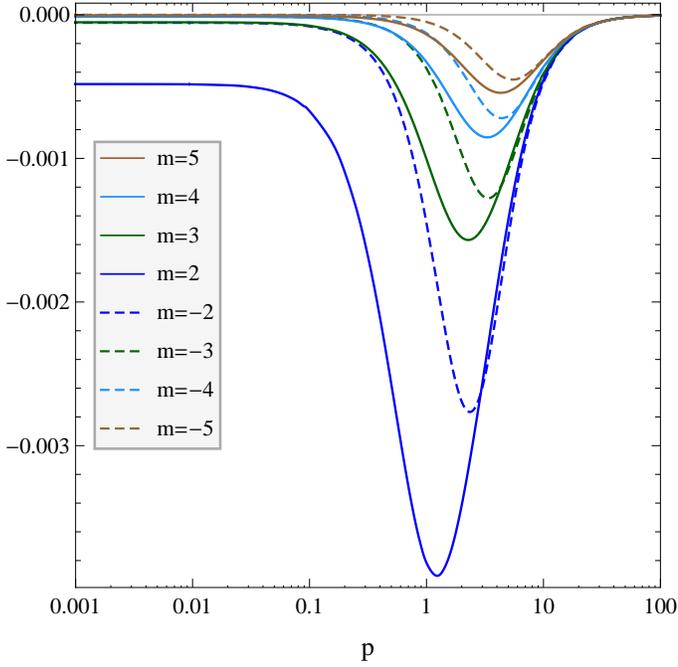}
\caption{\label{fig2}  Dependence of $\text{Re}[S_{1/2,1/2}^{\left(m\right)}-1]$
on the fermion momentum $p$ for $\vert m \vert = 2, 3, 4,\,\text{and}\; 5$.}
\end{figure}

\begin{figure}[tbp]
\includegraphics[width=0.5\textwidth]{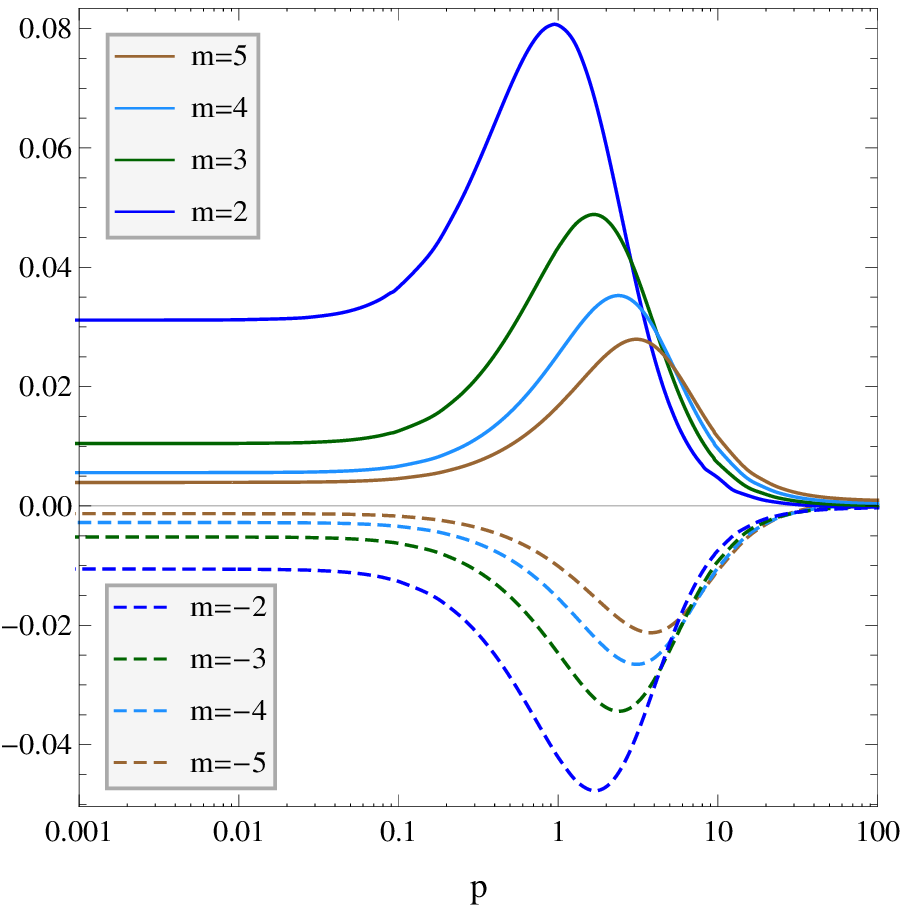}
\caption{\label{fig3}   Dependence of $\text{Im}[S_{1/2,1/2}^{\left(m\right)}]$
on the fermion momentum $p$ for $\vert m \vert = 2, 3, 4,\,\text{and}\; 5$.}
\end{figure}

\begin{figure}[tbp]
\includegraphics[width=0.5\textwidth]{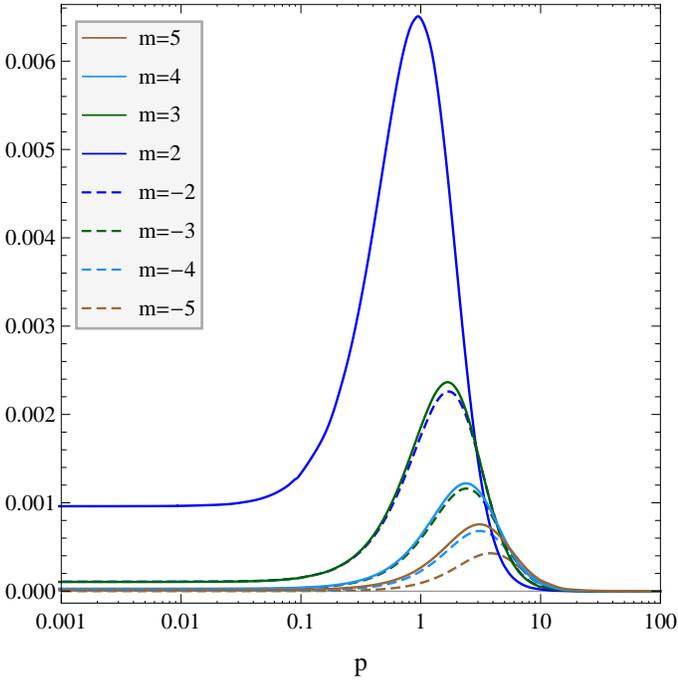}
\caption{\label{fig4}   Dependence of $\vert S_{1/2,1/2}^{\left( m \right)} - 1
\vert^{2}$ on the fermion momentum $p$ for $\vert m\vert=2,3,4,\,\text{and}\; 5
$.}
\end{figure}

\begin{figure}[tbp]
\includegraphics[width=0.5\textwidth]{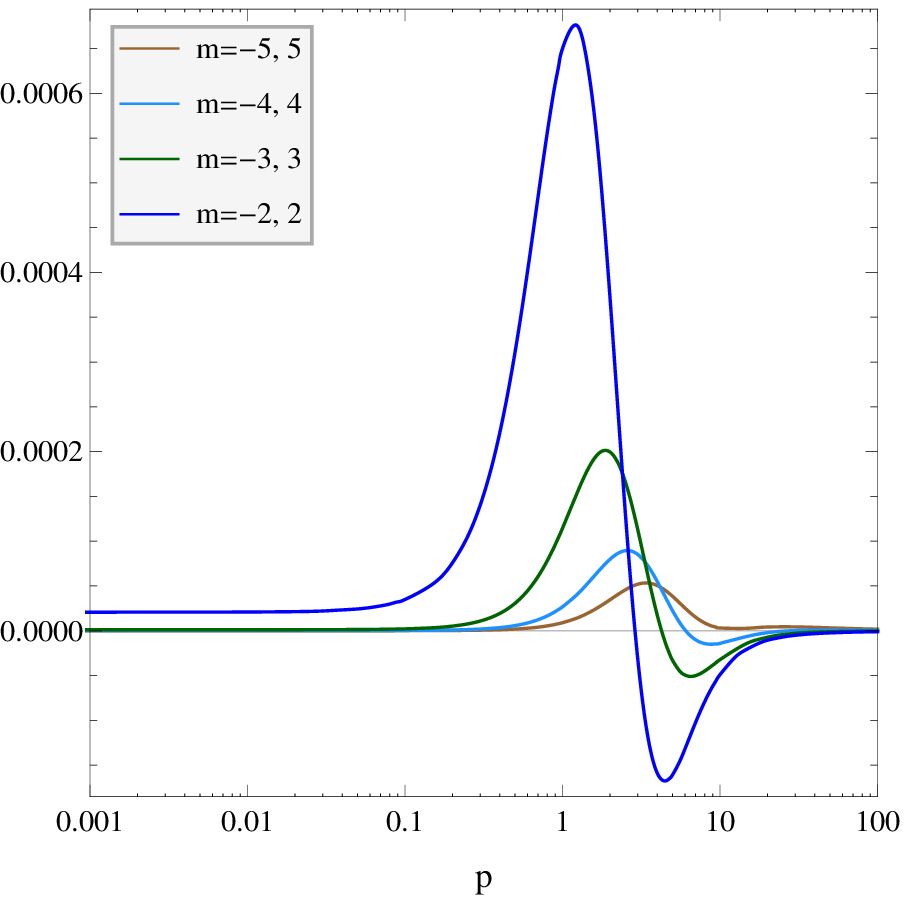}
\caption{\label{fig5}  Dependence of $\text{Re}[S_{-1/2,1/2}^{\left(m\right)}]$
on the fermion momentum $p$ for $\vert m \vert = 2, 3, 4,\,\text{and}\; 5$.}
\end{figure}

\begin{figure}[tbp]
\includegraphics[width=0.5\textwidth]{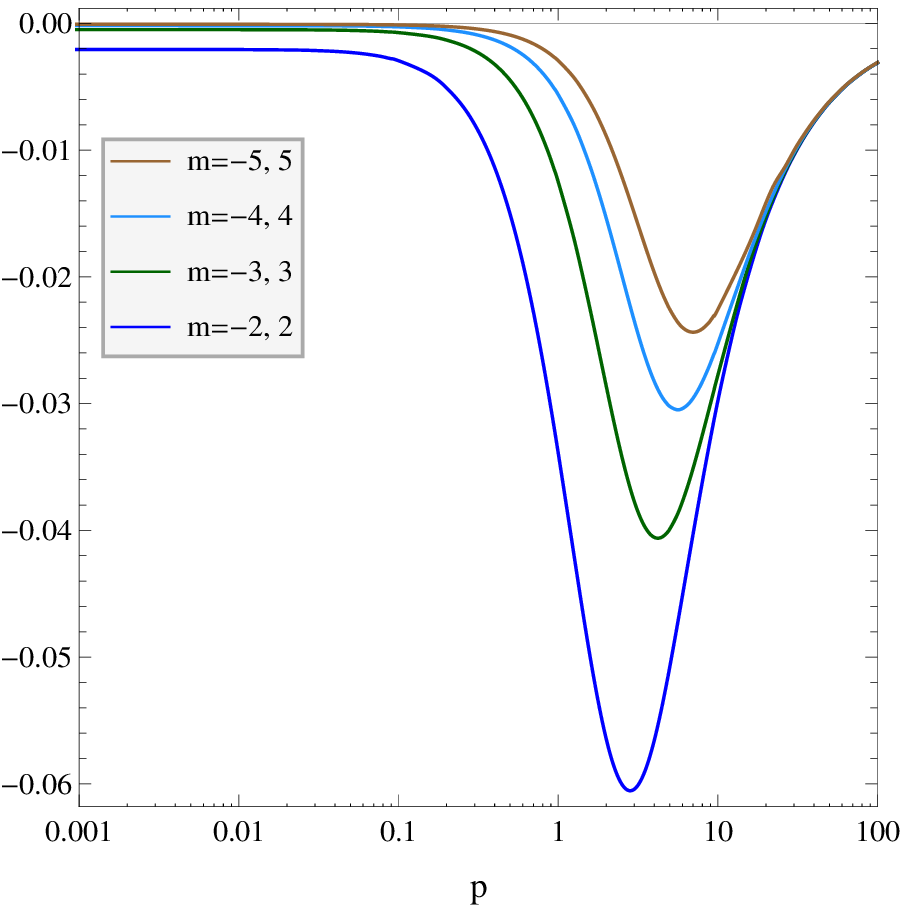}
\caption{\label{fig6}  Dependence of $\text{Im}[S_{-1/2,1/2}^{\left(m\right)}]$
on the fermion momentum $p$ for $\vert m \vert = 2, 3, 4,\,\text{and}\; 5$.}
\end{figure}

\begin{figure}[tbp]
\includegraphics[width=0.5\textwidth]{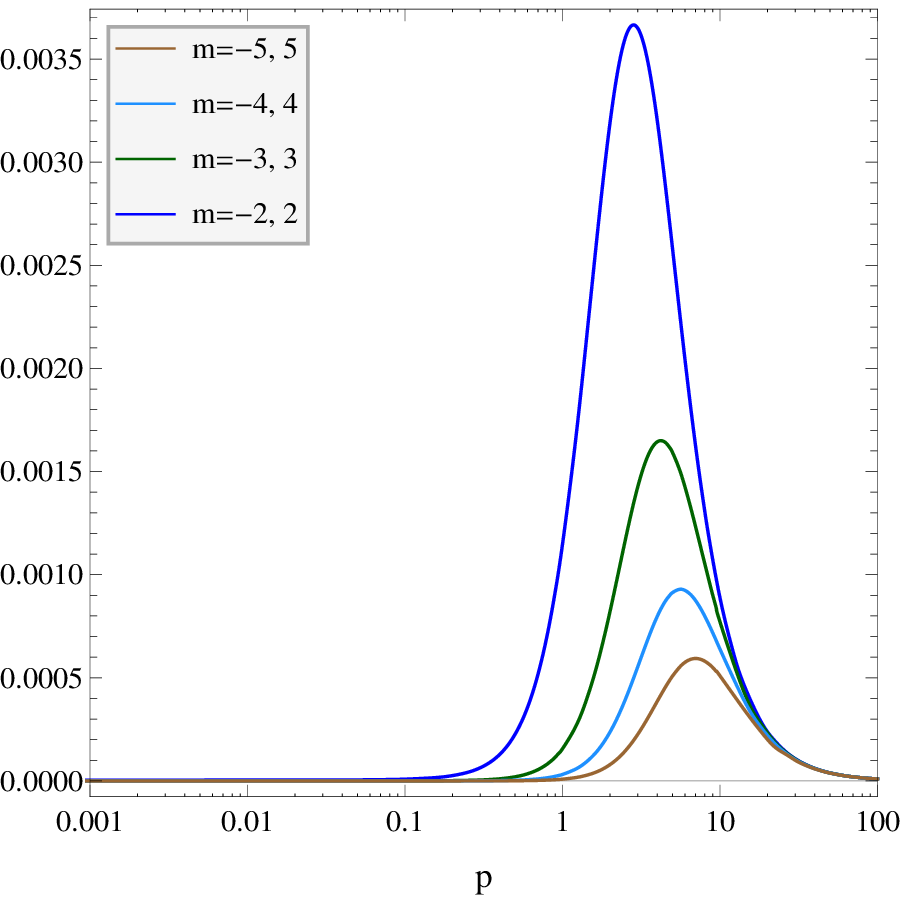}
\caption{\label{fig7}     Dependence of $\vert S_{-1/2, 1/2}^{\left( m \right)}
\vert^{2}$ on the fermion momentum $p$ for $\vert m\vert=2,3,4,\,\text{and}\; 5
$.}
\end{figure}

Figures~\ref{fig2} --  \ref{fig7}  show  the  real  parts, imaginary parts, and
squared magnitudes  of  the  partial matrix elements $S_{\pm 1/2,1/2}^{\left( m
\right)}$ as functions of the fermion momentum $p$.
The $p$-dependences are shown for $\left\vert m \right\vert  =  2, 3, 4, \text{
and}\,5$.
We can see that  for  $\left\vert m \right\vert \ge 2$, the curves of $\text{Re
}[S_{1/2,1/2}^{\left( m \right)}\left( p \right) -1]$, $\text{Im}[S_{1/2,1/2}^{
\left( m \right)}\left( p \right)]$, $\vert S_{1/2, 1/2}^{\left(m\right)}\left(
p\right)-1\vert^{2}$, $\text{Re}[S_{-1/2,1/2}^{\left(m\right)}\left(p\right)]$,
$\text{Im}[S_{-1/2,1/2}^{\left( m \right)}\left(p\right)]$, and $\vert S_{-1/2,
1/2}^{\left(m\right)}\left(p\right)\vert^{2}$ have similar forms.
All of them have wide extrema at moderate values of $p$, and tend to zero as $p
\rightarrow \infty$ and to constant values as $p \rightarrow 0$.
The only difference is  that for $\left\vert m \right\vert  =  2, 3, \text{and}
\,4$, the $\text{Re}[S_{-1/2,1/2}^{\left(m\right)} \left(p\right)]$ curves have
a node located to the right  of the maximum at moderate values of $p$.

It follows  from  Fig.~\ref{fig3}  that  the  imaginary  parts  of  the elastic
partial elements of the $S$-matrix satisfy the approximate relation
\begin{equation}
\text{Im}[S_{1/2,1/2}^{\left( m\right)}(p)]
\approx \text{Im}[S_{1/2,1/2}^{\left(-m + 1\right)}(p)].            \label{V:1}
\end{equation}
Since $\vert \text{Im}[S_{1/2,1/2}^{\left( m \right)}]\vert \gg \vert \text{Re}
[S_{1/2,1/2}^{\left( m \right)} - 1] \vert$,  approximate  equality (\ref{V:1})
results in the approximate equality of the squared magnitudes $\vert S_{1/2,1/2
}^{\left(m\right)}-1\vert^{2}\approx \vert S_{1/2,1/2}^{\left(-m + 1\right)} -1
\vert^{2}$, as shown in Fig.~\ref{fig4}.
It follows  from  Eqs.~(\ref{III:6})  and  (\ref{III:6a}) that for $n = 1$, the
change $m \rightarrow -m + 1$  of  the  grand  spin $K_{3}$  leads  only to the
interchange of the absolute values of the orbital angular momenta $\vert l_{11}
\vert$ and $\vert l_{21} \vert$.
It follows from Eq.~(\ref{III:9}) that the centrifugal  barrier does not change
as a  whole  for  the  elastic  channel  of  fermion  scattering,  and thus the
fermion-soliton interaction  is  of  the  same  order  for  the partial elastic
channels with $K_{3} = m$ and $K_{3} = -m + 1$.
This  may  explain  the  close  values  of  the  dominant  imaginary  parts  in
Eq.~(\ref{V:1}).

We also found that the position of the maximum  in  $\vert S_{1/2,1/2}^{\left(m
\right)}-1\vert^{2}$ is approximately determined by the linear expression
\begin{equation}
p_{\text{max}}\approx 0.64 \left\vert m - 1/2 \right\vert,          \label{V:3}
\end{equation}
whereas the heights of the maxima decrease monotonically (approximately $\propto
\left\vert m \right\vert^{-2}$) with an increase in $\left\vert m \right\vert$.
Note that Eq.~(\ref{V:3}) is  compatible with Eq.~(\ref{V:1}) since $\left\vert
m - 1/2 \right\vert$ is equal to $m-1/2$ for positive  $m$  and  $\left(- m + 1
\right) - 1/2$  for negative $m$.
Eq.~(\ref{V:3}) can be explained as follows.
The   partial  matrix  elements $S_{1/2, 1/2}^{\left( m \right)}$  describe the
elastic fermion scattering  in  the  state  with   grand  spin $K_{3} = m$.
Eq.~(\ref{III:4}) tells us that with an increase in $\left\vert m \right\vert$,
the main contribution to $K_{3}$ comes from the orbital part.
Next, it follows from Eqs.~(\ref{III:6})  and  (\ref{III:6a}) that for $n = 1$,
the orbital angular momenta  of  the  elastic  components $c_{11}$ and $c_{21}$
of  the  fermion  radial  wave  function are $l_{11} = m - 1$ and $l_{21} = m$,
respectively.
We see that both $l_{11}$ and $l_{21}$  depend linearly on $m$.
In the classical limit, the  absolute  value  of the  orbital  angular momentum
$\left\vert l \right\vert = b p$, where $b$ is an impact parameter.
In our case, the impact parameter  $b$  should be on the order of the soliton's
size $r_{0}$.
In the momentum representation, the fermion radial wave function $c_{i a}$ that
corresponds to  the  state  with  $K_{3} = m$  should  have  a  maximum  in the
neighborhood  of  the  classical  value  $p  =  \left\vert l \right\vert/b$, in
accordance with Eq.~(\ref{V:3}).

It  follows  from   Figs.~\ref{fig5} -- \ref{fig7}   that  in  accordance  with
Eq.~(\ref{III:25}), the  inelastic   partial  matrix  elements $S_{-1/2, 1/2}^{
\left( m \right)}$ coincide when they have opposite values of $m$.
Furthermore, similarly   to   Eq.~(\ref{V:3}) and  for  the  same  reasons, the
position of the  maximum  of  $\vert S_{-1/2,1/2}^{\left( m \right)} \vert^{2}$
is also determined by the linear expression
\begin{equation}
p_{\text{max}}\approx 1.37\left\vert m \right\vert.                 \label{V:4}
\end{equation}

From a comparison between the  $p$-dependences  related  to  $S_{\pm 1/2,1/2}^{
\left(0\right)}$ and those  related  to  $S_{\pm 1/2, 1/2}^{\left(\vert m \vert
\ge 2\right)}$, we can see two main differences.
Firstly, there are no pronounced  maxima  for $\vert S_{I_{3}^{\prime},I_{3}}^{
\left(0\right)}-\delta_{I_{3}^{\prime},I_{3}}\vert^{2}$  at  nonzero $p$, since
for $m = 0$, the  contribution  of  the orbital part  to the grand spin $K_{3}$
is not  dominant  in  comparison  with  those  of  the  spin  and isospin parts.
Secondly,  the  limiting  values  of  $\vert  S_{I_{3}^{\prime},I_{3}}^{\left(0
\right)} - \delta_{I_{3}^{\prime },I_{3}}\vert^{2}$ are much greater than those
of $\vert S_{I_{3}^{\prime},I_{3}}^{\left(\left\vert m\right \vert > 2 \right)}
-\delta_{I_{3}^{\prime},I_{3}}\vert ^{2}$ as $p \rightarrow 0$.
This is  because  according  to  Eq.~(\ref{III:6a}), both $l_{21}$ and $l_{12}$
vanish when $m=0$ and $n=1$.
It follows that  in  this  case,  the  centrifugal  barrier  is absent for both
elastic and inelastic fermion scattering.
This leads to intense interactions  between  the  fermions  and the core of the
soliton.
In turn, this  results  in  large  squared  magnitudes for both the elastic and
inelastic partial elements of the $S$-matrix.

Using numerical methods, we  were also able to ascertain some other features of
the curves shown in Figs.~\ref{fig1} -- \ref{fig7}.
In particular, we ascertained the asymptotic behavior of $S_{I_{3}^{\prime},I_{
3}}^{\left(m\right)}(p)$ as $p \rightarrow \infty$:
\begin{subequations}                                                \label{V:5}
\begin{eqnarray}
\text{Re}\left[ S_{1/2,1/2}^{\left( m\right) }-1\right]  &\sim &\alpha
\left( m\right) p^{-2},                                            \label{V:5a}
\\
\text{Im}\left[ S_{1/2,1/2}^{\left( m\right) }\right]  &\sim &\beta \left(
m\right) p^{-1},                                                   \label{V:5b}
\\
\text{Re}\left[ S_{-1/2,1/2}^{\left( m\right) }\right]  &\sim &\gamma \left(
m\right) p^{-1},                                                   \label{V:5c}
\\
\text{Im}\left[ S_{-1/2,1/2}^{\left( m\right) }\right]  &\sim &-\pi h p^{-1},
                                                                   \label{V:5d}
\end{eqnarray}
\end{subequations}
where $\alpha\left(m\right)$, $\beta\left(m\right)$, and $\gamma\left(m\right)$
are $m$-dependent constants.
Note that for $\text{Im}[S_{-1/2,1/2}^{\left( m\right)}]$, we were able to find
the exact form of the coefficient of  the leading asymptotic term.
This is because the Born  approximation (\ref{IV:10'b}) perfectly describes the
behavior  of  $\text{Im}[S_{-1/2, 1/2}^{\left( m \right)}]$ for $p \gtrsim 10$.
At the same time, the Born approximation (\ref{IV:10'a}) gives only a qualitative
description  of  $\text{Im}[S_{1/2,1/2}^{\left(m\right)}]$: it gives the correct
($\propto p^{-1}$) leading asymptotic  behavior, but an incorrect  factor before
the leading asymptotic term.
Note that in Eq.~(\ref{V:5a}), the leading asymptotic term is $\propto p^{-2}$,
while  in   Eqs.~(\ref{V:5b}) -- (\ref{V:5d}),  the  leading  asymptotic  terms
are $\propto p^{-1}$.
This difference is due to the fact that $\text{Im}[ S_{1/2,1/2}^{\left(m\right)
}]$, $\text{Re}[S_{-1/2,1/2}^{\left(m\right)}]$, and $\text{Im}[ S_{-1/2,1/2}^{
\left( m \right)}]$  tend  to  zero as $p \rightarrow \infty$, while $\text{Re}
[S_{1/2,1/2}^{\left(m\right)}]$ tends to one in this limit.
It can then easily be shown that asymptotic  behavior (\ref{V:5a}) follows from
the fulfillment of the unitarity  condition $\vert S_{1/2,1/2}^{\left(m\right)}
\vert ^{2} + \vert S_{-1/2,1/2}^{\left( m\right) }\vert^{2} = 1$ in the leading
order in inverse powers of $p$.

It  follows  from  Figs.~\ref{fig1} --  \ref{fig7}  that  as $p$ tends to zero,
the  real  and  imaginary  parts of the difference $S_{I_{3}', I_{3}}^{\left( m
\right)}-\delta_{I_{3}', I_{3}}$ tend to some constants  whose  absolute values
 decrease monotonically  with an increase in $\left\vert m \right\vert$.
Then,  Eqs.~(\ref{III:17})   and  (\ref{III:18'}) tell us that both the elastic
and inelastic  partial cross-sections  $\sigma_{I_{3}^{\prime},I_{3}}^{\left( m
\right)} = p^{-1}\vert S_{I_{3}^{\prime},I_{3}}^{\left(m\right)} - \delta_{I_{3
}^{\prime},I_{3}}\vert^{2}$ diverge as $p^{-1}$ when $p \rightarrow 0$.

Next we turn to  the   partial  matrix  elements  $S_{\pm 1/2, 1/2}^{\left( \pm
1 \right)}$.
The dependence of  the  real  and  imaginary  parts of $S_{\pm 1/2,1/2}^{\left(
+ 1\right)}$ on the fermion momentum $p$ is shown in Fig.~\ref{fig8}.
We can see that the $p$-dependences in Fig.~\ref{fig8} are in sharp contrast to
those in Figs.~\ref{fig1} -- \ref{fig7}.
In particular, the $p$-dependence  of the elastic  partial  element $S_{1/2,1/2
}^{\left( +1 \right)}$ shows  pronounced  resonance behavior  at  extremely low
values of $p$.
We were able to achieve extremely small  values  of $p$ to reveal two resonance
valleys  of  $\text{Re}[ S_{1/2,1/2}^{\left(+1\right)}]$.
The positions of  the  extrema  in  $\text{Re}[ S_{1/2,1/2}^{\left(+1\right)}]$
coincide with those of the zeros of $\text{Im}[S_{1/2,1/2}^{\left(+1\right)}]$,
as expected for resonance structures.
Note that $\text{Re}[S_{1/2,1/2}^{\left(+1\right)}]$ (and  hence $\vert S_{1/2,
1/2}^{\left( +1 \right)}\vert$) reaches  the  unitary  boundary  of $1$ at its
maxima.
Hence, the elastic partial  cross-section $\sigma_{1/2,1/2}^{(+1)}$ vanishes at
the  maxima  of  $\text{Re}[S_{1/2,1/2}^{\left( +1 \right)}]$  as  well  as the
inelastic partial cross-section $\sigma_{-1/2,1/2}^{(+1)}$.
Note that both $\text{Re}[S_{1/2, 1/2}^{\left( +1 \right)}]$ and $\text{Im}[S_{
1/2,1/2}^{\left(+1\right)}]$ do not reach the unitary boundary of $-1$ at their
minima, although these are in close proximity.
This follows from the fact that the inelastic scattering does not vanish at the
minima of $\text{Re}[S_{1/2,1/2}^{\left( +1\right)}]$ and $\text{Im}[S_{1/2,1/2
}^{\left(+1 \right)}]$.
For a similar reason, $\text{Im}[S_{1/2,1/2}^{\left(+1\right)}]$  also does not
quite reach the unitary boundary of $1$ at its maxima.

The $p$-dependences of the real and imaginary parts  of  the inelastic  partial
matrix element $S_{-1/2,1/2}^{\left(+1\right)}$ also have a resonance structure
at small values of $p$.
Moreover, the  positions  of  the  minima (maxima) of $\text{Re}[S_{-1/2,1/2}^{
\left(+1\right)}]$ coincide with those of  the  maxima (minima) of  $\text{Re}[
S_{1/2,1/2}^{\left(+1\right)}]$, and a similar situation holds for the imaginary
parts $\text{Im}[S_{-1/2,1/2}^{\left(+1\right)}]$  and $\text{Im}[S_{1/2,1/2}^{
\left(+1\right)}]$.
The positions  of  the  zeros  of  $\text{Re}[S_{-1/2,1/2}^{\left(+1\right)}]$,
$\text{Im}[S_{-1/2,1/2}^{\left(+1\right)}]$, and $\text{Im}[S_{1/2,1/2}^{\left(
+1 \right)}]$ also coincide, since  $\text{Re}[S_{1/2,1/2}^{\left( +1\right)}]$
reaches the unitary boundary of $1$ at this point.
Note that  in  Fig.~\ref{fig8}, the  behavior  of  the  curves is in accordance
with the unitarity  condition  $\vert S_{1/2,1/2}^{\left( +1 \right)}\vert^{2}+
\vert S_{-1/2,1/2}^{\left( +1\right) }\vert^{2} = 1$.

Figure~\ref{fig9}  presents the $p$-dependence of the real and imaginary parts
of the  partial matrix elements  $S_{\pm 1/2,1/2}^{\left(-1\right)}$.
We see that in  accordance  with  Eq.~(\ref{III:25}), the curves of $\text{Re}[
S_{-1/2,1/2}^{\left(-1\right)}]$  and $\text{Im}[S_{-1/2,1/2}^{\left(-1\right)}
]$ coincide with the curves for $\text{Re}[S_{-1/2,1/2}^{\left(+1\right)}]$ and
$\text{Im}[S_{-1/2,1/2}^{\left(+1\right)}]$ shown in Fig.~\ref{fig8}.
In particular, the curves  of  $\text{Re}[S_{-1/2,1/2}^{\left(-1\right)}]$  and
$\text{Im}[S_{-1/2,1/2}^{\left(-1\right)}]$  have  the same resonance structure
as those of $\text{Re}[S_{-1/2,1/2}^{\left(+1\right)}]$ and $\text{Im}[S_{-1/2,
1/2}^{\left(+1\right)}]$, respectively.
In Fig.~\ref{fig9}, the  curves that correspond to the real and imaginary parts
of the elastic partial matrix element $S_{1/2,1/2}^{\left(-1\right)}$ also show
resonance-like behavior.
However,  the  behavior  of  these  curves  differs  from  those  of  the  other
resonance curves in Figs.~\ref{fig8} and \ref{fig9}.
In particular, the points of the maxima in $\vert S_{1/2,1/2}^{\left(-1\right)}
\vert^{2}$ and $\text{Re}[S_{1/2,1/2}^{\left(-1\right)}]$  do not coincide, and
$\text{Im}[S_{1/2,1/2}^{\left( -1\right)}]$  does  not  vanish at these points.
Hence, it can be said that $S_{1/2,1/2}^{\left( -1 \right)}$ does not show true
resonance behavior.
Instead, the  resonance-like   behavior  of  $S_{1/2,1/2}^{\left(-1\right)}$ is
caused by the unitarity condition $\vert S_{1/2,1/2}^{\left(-1\right)}\vert^{2}
+\vert S_{-1/2,1/2}^{\left( -1 \right) }\vert^{2} = 1$  and  the true resonance
behavior of the  inelastic  partial element $S_{-1/2,1/2}^{\left(-1\right)}$ of
the $S$-matrix.

In  Figs.~\ref{fig8}  and  \ref{fig9},  all  $p$-dependencies  are  shown  on a
logarithmic scale.
We see that in the resonance region, the $p$-dependences  of $\vert S_{\pm 1/2,
1/2}^{\left(\pm 1\right)}-\delta_{\pm 1/2,1/2}\vert^{2}$, $\text{Re}[S_{\pm 1/2,
1/2}^{\left(\pm 1 \right)}-\delta_{\pm 1/2, 1/2}]$, and  $\text{Im}[S_{\pm 1/2,
1/2}^{\left(\pm 1\right)}]$ have the Breit-Wigner form on this scale.
Also, the positions  of  resonance  peaks  are  at  extremely low values of the
fermion momenta.
Another feature is a periodic structure  of  the resonances in Figs.~\ref{fig8}
and  \ref{fig9}.
In particular, the distance  between  the  resonance peaks is approximately the
same on the logarithmic scale.
We were able to identify the two  resonance  peaks of $\vert S_{\pm 1/2, 1/2}^{
\left(\pm 1\right)} - \delta_{\pm 1/2,1/2}\vert^{2}$ and reach the beginning of
the third one.
Therefore, we may suppose the existence of  a  sequence of resonances (possibly
infinite) condensing to the zero fermion momentum.

\section{\label{sec:VI} Conclusion}

In the  present  paper,   fermion  scattering  in the background fields of the
topological  solitons   of   the   nonlinear  $O(3)$  $\sigma$-model  has  been
investigated both analytically and numerically.
In particular,  we have ascertained the asymptotic behavior of the fermion wave
functions for both large and small values of the radial variable $r$.
The symmetry properties of the partial elements of the $S$-matrix under discrete
transformations of the Dirac Hamiltonian have been determined.
In the  framework  of  the  first  Born  approximation,  a  complete analytical
investigation of fermion scattering in  the  background field of the elementary
topological soliton with  winding  number $n = 1$ has been carried out.
In particular, the  Born  amplitudes,   differential  cross-sections, and total
cross-sections of fermion scattering  have  been  obtained  in analytical form.
The asymptotic behavior of  the partial Born  amplitudes  has been investigated
for extreme values of  the fermion momentum, momentum transfer, and grand spin.

We have also performed a numerical study of fermion scattering in the background
field  of  the  elementary  topological  solition  of   the   nonlinear  $O(3)$
$\sigma$-model.
In particular, we have obtained  the  $p$-dependences  of the  partial elements
$S_{I_{3}',I_{3}}^{\left( m \right)}$   of  the  $S$-matrix  for  $\left\vert m
\right\vert\le 5$, and have ascertained their main features.
In particular, we found that the  partial elements $S_{I_{3}',I_{3}}^{\left(\pm
1 \right)}$ of the $S$-matrix show  resonance  behavior  at small values of the
fermion momentum.

\begin{acknowledgments}

This work was supported by the Russian Science Foundation, grant No 19-11-00005.

\end{acknowledgments}

\appendix

\section{Free fermions}

Let the isovector  scalar  field  $\boldsymbol{\phi}$  be a constant field that
takes the value $(0, 0, -1)$.
This situation corresponds to  the  vacuum  state  in the topologically trivial
sector $(n = 0)$ or distant  regions  in  the  topologically nontrivial sectors
$(n \ne 0)$.
In this case, the Dirac equation (\ref{II:5b}) is written as
\begin{equation}
i\gamma^{\mu}\otimes\mathbb{I}\partial_{\mu}\psi - h \mathbb{I} \otimes
\tau_{3}\psi = 0                                                    \label{A:1}
\end{equation}
or in the Hamiltonian form:
\begin{equation}
i \frac{\partial \psi }{\partial t} = H_{0} \psi,                   \label{A:2}
\end{equation}
where the free Hamiltonian
\begin{equation}
H_{0} = \alpha^{k}\otimes\mathbb{I}\left( -i \partial_{k} \right)
+ h \beta \otimes \tau_{3}.                                         \label{A:3}
\end{equation}
The Dirac equation (\ref{A:1}) is invariant under  the  $C$, $P$ and $\Pi_{2}T$
transformations:
\begin{eqnarray}
\psi \left( t,\mathbf{x}\right)  &\rightarrow &\psi ^{C}\left( t,\mathbf{x}
\right) =i\gamma ^{1}\mathbb{\otimes I\,}\psi ^{\ast }\left( t,\mathbf{x}
\right),                                                            \label{A:4}
\\
\psi \left( t,\mathbf{x}\right) &\rightarrow &\psi ^{P}\left( t,\mathbf{x}
\right) = \gamma^{0}\mathbb{\otimes I}
\psi \left( t,-\mathbf{x}\right),                                   \label{A:5}
\\
\psi \left(t,\mathbf{x}\right) &\rightarrow & \psi^{\Pi_{2}T}\left(t,\mathbf{x}
\right) = \psi^{\ast}\left(-t, x_{1}, -x_{2}\right).               \label{A:5'}
\end{eqnarray}
The  Hamiltonian  (\ref{A:3})  commutes   with  the  operator   of  grand  spin
(\ref{III:4}), which is reduced to the operator  of  the usual angular momentum
in the topologically trivial background vacuum field $\boldsymbol{\phi}_{\text{
vac}} = (0, 0, -1)$.
It also commutes with the isospin generator $I_{3}=\tau_{3}/2$ and the momentum
operator $\mathbf{\hat{p}}=-i\nabla$.
It therefore follows that the free fermionic states can be characterized either
by the momentum  $\mathbf{p}$ and the  third  isospin  component $I_{3}$ (plane
waves) or by the grand spin $K_{3}$ and the  third  isospin  component  $I_{3}$
(cylindrical waves).

In the compact matrix form,  the  plane-wave fermionic states $\psi_{\pm p, I_{
3}}$ with positive and negative energies can be written as
\begin{subequations}                                                \label{A:6}
\begin{equation}
\psi _{p,1/2}=\frac{1}{\sqrt{2\varepsilon }}
\begin{pmatrix}
\sqrt{\varepsilon +h} & 0 \\
i\sqrt{\varepsilon -h}e^{i\theta _{\mathbf{p}}} & 0
\end{pmatrix}
e^{-ipx},                                                          \label{A:6a}
\end{equation}
\begin{equation}
\psi _{p,-1/2}=\frac{1}{\sqrt{2\varepsilon }}
\begin{pmatrix}
0 & -i\sqrt{\varepsilon -h}e^{-i\theta _{\mathbf{p}}} \\
0 & \sqrt{\varepsilon +h}
\end{pmatrix}
e^{-ipx},                                                          \label{A:6b}
\end{equation}
\begin{equation}
\psi _{-p,1/2}=\frac{1}{\sqrt{2\varepsilon }}
\begin{pmatrix}
-i\sqrt{\varepsilon -h}e^{-i\theta _{\mathbf{p}}} & 0 \\
\sqrt{\varepsilon +h} & 0
\end{pmatrix}
e^{ipx},                                                           \label{A:6c}
\end{equation}
\begin{equation}
\psi _{-p,-1/2}=\frac{1}{\sqrt{2\varepsilon }}
\begin{pmatrix}
0 & \sqrt{\varepsilon + h} \\
0 & i\sqrt{\varepsilon - h}e^{i\theta_{\mathbf{p}}}
\end{pmatrix}
e^{ipx},                                                           \label{A:6d}
\end{equation}
\end{subequations}
where $p=\left(\varepsilon, \mathbf{p}\right)$, $px = \varepsilon t -\mathbf{p}
\!\boldsymbol{\cdot}\!\mathbf{x}$, and $\theta_{\mathbf{p}}$  is  the azimuthal
angle of the momentum $\mathbf{p}$.
Note that the negative energy wave functions are $C$-conjugates of the positive
energy wave functions, $\psi_{-p,\,I_{3}} = \left(\psi_{p,\,I_{3}}\right)^{C} =
\sigma_{1} \mathbb{\otimes I\,}\psi _{p,\,I_{3}}^{\ast }$.
The wave functions $\psi_{\pm p,I_{3}}$  and their amplitudes $u_{\pm p,I_{3}}$
defined by the formula $\psi_{\pm p,I_{3}}=\left(2\varepsilon \right)^{-1/2}u_{
\pm p,I_{3}}e^{\mp i p x}$ satisfy the normalization conditions:
\begin{equation}
\bar{\psi}_{\pm p,\,I_{3}}\gamma^{\mu}\otimes \mathbb{I} \psi_{\pm p,\,I_{3}}
=\left(1,\frac{p_{x}}{\varepsilon },\frac{p_{y}}{\varepsilon }\right)=
\left(1,\,\mathbf{v}\right)                                         \label{A:7}
\end{equation}
and
\begin{equation}
\bar{u}_{\pm p,I_{3}^{\prime }}u_{\pm p,\,I_{3}}=(-1)^{1/2\mp
I_{3}} 2 h \delta _{I_{3}^{\prime},I_{3}},\quad
\bar{u}_{p,I_{3}^{\prime}} u_{-p,\,I_{3}} = 0.                      \label{A:8}
\end{equation}
It follows that the wave  functions $\psi_{\pm p,I_{3}}$ have the normalization
\begin{align}
\int \bar{\psi }_{\pm p^{\prime },I_{3}^{\prime }}(\mathbf{x})\psi
_{\pm p,I_{3}}(\mathbf{x})d^{2}\mathbf{x} =&(-1)^{1/2\mp I_{3}}\frac{h}{
\varepsilon }\left( 2\pi \right) ^{2}                            \nonumber
\\
& \times \delta _{I_{3}^{\prime}, I_{3}}\delta ^{\left( 2\right) }\left(
\mathbf{p}-\mathbf{p}^{\prime }\right),                            \label{A:8b}
\end{align}
where it is  understood  that in Eqs.~(\ref{A:7})--(\ref{A:8b}),  the summation
is performed  over  the  spin  and  isospin  indices  of the corresponding wave
functions and amplitudes.

Let us consider  the  cylinder-wave  fermionic  states  $\psi_{p,m,I_{3}}$ that
possess definite values of  the grand  spin $K_{3}$  and  the  isospin $I_{3}$.
For free fermions, the  matrix  $\Lambda$  in  Eq.~(\ref{III:7}) takes the form
\begin{equation}
\Lambda_{i a; j b} = \left(
\begin{array}{cccc}
\dfrac{l_{11}}{r} & 0 & \varepsilon +h & 0 \\
0 & \dfrac{l_{12}}{r} & 0 & \varepsilon -h \\
-\varepsilon +h & 0 & -\dfrac{l_{21}}{r} & 0 \\
0 & -\varepsilon -h & 0 & -\dfrac{l_{22}}{r}
\end{array}
\right),                                                            \label{A:9}
\end{equation}
where the orbital quantum  numbers $l_{i a}$ are defined in Eq.~(\ref{III:6a}).
We see that system (\ref{III:7}) can be split  into  two independent subsystems
that correspond to the two isospin states $\pm 1/2$.
These two subsystems can  be  solved  analytically,  and  their positive energy
regular solutions are
\begin{widetext}
\begin{subequations}                                               \label{A:10}
\begin{flalign}
\psi _{p,m,1/2}& =
\begin{pmatrix}
\sqrt{\frac{p\left( \varepsilon +h\right) }{2\varepsilon }}J_{l_{11}}\left(
pr\right) e^{il_{11}\theta } & 0 \\
-\sqrt{\frac{p\left( \varepsilon -h\right) }{2\varepsilon }}J_{l_{21}}\left(
pr\right) e^{il_{21}\theta } & 0%
\end{pmatrix}%
e^{-i\varepsilon t}\sim \sqrt{\frac{2}{\pi r}}%
\begin{pmatrix}
\sqrt{\frac{\varepsilon +h}{2\varepsilon }}\sin \left[ pr-\pi \left(
2l_{11}-1\right) /4\right] e^{il_{11}\theta } & 0 \\
\sqrt{\frac{\varepsilon -h}{2\varepsilon }}\sin \left[ pr-\pi \left(
2l_{21}+3\right) /4\right] e^{il_{21}\theta } & 0%
\end{pmatrix} e^{-i\varepsilon t},                                \label{A:10a}
\end{flalign}
\begin{flalign}
\psi _{p,m,-1/2}& =
\begin{pmatrix}
0 & -\sqrt{\frac{p\left( \varepsilon -h\right) }{2\varepsilon }}%
J_{l_{12}}\left( pr\right) e^{il_{12}\theta } \\
0 & \sqrt{\frac{p\left( \varepsilon +h\right) }{2\varepsilon }}%
J_{l_{22}}\left( pr\right) e^{il_{22}\theta }%
\end{pmatrix}%
e^{-i\varepsilon t}\sim \sqrt{\frac{2}{\pi r}}%
\begin{pmatrix}
0 & \sqrt{\frac{\varepsilon -h}{2\varepsilon }}\sin \left[ pr-\pi \left(
2l_{12}+3\right) /4\right] e^{il_{12}\theta } \\
0 & \sqrt{\frac{\varepsilon +h}{2\varepsilon }}\sin \left[ pr-\pi \left(
2l_{22}-1\right) /4\right] e^{il_{22}\theta }%
\end{pmatrix} e^{-i\varepsilon t},                                \label{A:10b}
\end{flalign}
\end{subequations}
\end{widetext}
where $p = \left\vert \bf{p} \right\vert$  and  $J_{\nu}(p r)$  are  the Bessel
functions of the first kind.
The negative energy regular solutions  are  obtained  from Eqs.~(\ref{A:10}) by
means of $C$-conjugation (\ref{A:4}).
Wave   functions    (\ref{A:10})    satisfy    the    normalization   condition
\begin{align}
\int \psi _{p^{\prime },m^{\prime },I_{3}^{\prime }}^{\ast }(r,\theta )\psi
_{p,m,I_{3}}(r,\theta )rdrd\theta  =&2\pi \delta_{m^{\prime}\!,m}\delta
_{I_{3}^{\prime}, I_{3}}                                          \nonumber
\\
&\times \delta \left(p\!-\!p^{\prime }\right)\!.                   \label{A:11}
\end{align}
%where  the  summation  is  performed  over the  spin and isospin indices of the
%wave functions.
Unlike wave  functions (\ref{A:6}),  wave  functions (\ref{A:10}) have definite
parity under $P$-transformation (\ref{A:5})
\begin{align}
\psi^{P}_{p,m,I_{3}}\left( t,\mathbf{x}\right)  &=\gamma
^{0}\!\otimes\!\mathbb{I}\psi_{p,m,I_{3}}\left(t,-\mathbf{x}\right) \nonumber
\\
&=\left( -1\right) ^{m-nI_{3}-\frac{1}{2}}\psi_{p,m,I_{3}}\left( t, \mathbf{
x}\right)\!                                                        \label{A:12}
\end{align}
and are invariant under $\Pi_{2}T$-transformation (\ref{A:5'}).

The plane-wave fermionic  states  $\psi_{p,I_{3}}$ can  be expanded in terms of
the cylinder-wave fermionic states $\psi_{p,m,I_{3}}$.
To do this, we use the  well-known  expansion  of  the  plane  wave in terms of
cylinder waves
\begin{equation}
e^{i p x}=e^{i p r\cos \left( \theta \right) }=
\sum_{i=-\infty}^{\infty}i^{m}J_{m}\left(p r\right) e^{i m\theta}. \label{A:13}
\end{equation}
Eqs.~(\ref{A:6}),  (\ref{A:10}),   and   (\ref{A:13})  give  us  the  expansion
\begin{equation}
\psi_{\left( \varepsilon ,p,0\right),\,I_{3}}=\sum_{m=-\infty }^{\infty
}a_{p,m,I_{3}}\psi_{p,m,I_{3}},                                    \label{A:14}
\end{equation}
where the expansion coefficients
\begin{equation}
a_{p,m,I_{3}}=i^{m-I_{3}\left( n+1\right) }p^{-1/2}                \label{A:15}
\end{equation}
and all components of the $(2+1)$-dimensional  momentum of the plane-wave state
are explicitly shown on the left-hand side of Eq.~(\ref{A:14}).
Note that Eq.~(\ref{A:15}) is  valid  only  for  the  positive energy fermionic
states.
To obtain the expansion coefficients for  the negative energy fermionic states,
we must take the complex conjugate of the right-hand  side of Eq.~(\ref{A:15}).

\section{Resonances of partial amplitudes at small fermion momenta}

Let us ascertain the cause of occurrence of  the resonance peaks in the partial
channel with grand spin  $m = 1$. % and  their logarithmic character.
It  follows  from  Eqs.~(\ref{III:6a}), (\ref{III:9}), and (\ref{III:17'}) that
the characteristic feature of this  partial  channel  is  the  absence  of both
the  kinematic suppression factor $(\varepsilon - h)^{1/2}$ and the centrifugal
barrier for  the  $c_{11}$  component of the radial wave function $c_{i a}(r)$.
As a result, the  $c_{11}$  component  is  much  larger  than  the  other three
components of $c_{i a}(r)$.
Using this fact, we can find  an  approximate  solution to system (\ref{III:7})
for small values of $p$.
To do this, we perform two iterations.
At the first iteration,  we suppose that $c_{11}(r)$ is a constant $\alpha_{0}$
and find $c_{12}(r)$, $c_{21}(r)$, and $c_{22}(r)$ neglecting  all terms except
centrifugal and those that proportional to $\alpha_{0}$.
At  the   second  iteration,   we   substitute  $c_{12}(r)$,  $c_{21}(r)$,  and
$c_{22}(r)$ into the differential equation  for  $c_{11}(r)$ and integrate this
equation.
As a result, we obtain the  approximate  iterative solution for the radial wave
function $c_{i a}(r)$
\begin{subequations}                                                \label{B:1}
\begin{eqnarray}
c_{11}\left( r\right)  &=&\frac{\alpha _{0}}{12}\Bigl\{
12-3p^{2}r^{2}-hr_{0}^{2}\left( \varepsilon +h\right) \Bigr. \nonumber \\
&&\times \Bigl( \pi ^{2}-3\ln \left[ \left( r^{2}+r_{0}^{2}\right) r_{0}^{-2}%
\right] \Bigr. \nonumber \\
&&\times \left. \left( 2+\ln \left[ r_{0}^{2}\left( r^{2}+r_{0}^{2}\right)
r^{-4}\right] \right) \right. \nonumber \\
&&\left. \left. -6\text{Li}_{2}\left[ r_{0}^{2}\left( r^{2}+r_{0}^{2}\right)
^{-1}\right] \right) \right\}, \\
c_{12}\left( r\right)  &=&\beta _{0}r-2^{-1}\alpha_{0}hr_{0}^{3}\left(
\varepsilon +h\right) r^{-1} \nonumber \\
&&\times \ln \left[ \left( r^{2}+r_{0}^{2}\right) r_{0}^{-2}\right] , \\
c_{21}\left( r\right)  &=&-2^{-1}\alpha _{0}\left( \varepsilon -h\right)
r-\alpha _{0}hr_{0}^{2}r^{-1} \nonumber \\
&&\times \ln \left[ \left( r^{2}+r_{0}^{2}\right) r_{0}^{-2}\right] , \\
c_{22}\left( r\right)  &=&\alpha _{0}hr_{0}\left(r_{0}^{2}r^{-2}\ln
\left[\left(r^{2}\!+\!r_{0}^{2}\right)r_{0}^{-2}\right]\!-\!1\right),
\end{eqnarray}
\end{subequations}
where $\alpha_{0}$ and  $\beta_{0}$  are  constants, and $\text{Li}_{2}$ is the
dilogarithm function.

At small $r$, the components  of  the  radial  wave  function can be written as
\begin{subequations}                                                \label{B:2}
\begin{eqnarray}
c_{11}\left( r\right)  & = &\alpha _{0}-2^{-2}\alpha _{0}p^{2}r^{2}
+ O\left(r^{4}\right), \\
c_{12}\left( r\right)  & = &\left(\beta_{0} - 2^{-1}\alpha
_{0}hr_{0}\left( \varepsilon+h\right)\right)r \nonumber \\
& & + O\left(r^{3}\right)\!, \\
c_{21}\left( r\right)  & = &-2^{-1}\alpha _{0}\left( \varepsilon +h\right)
r + O\left(r^{3}\right), \\
c_{22}\left( r\right)  & = &-2^{-1}\alpha _{0}hr_{0}^{-1}r^{2}
+ O\left(r^{4}\right)
\end{eqnarray}
\end{subequations}
in accordance with Eq.~(\ref{III:9}).
It was found that in a wide area of $r$, the  components  of  the  radial  wave
function satisfy the conditions
\begin{subequations}
\begin{eqnarray}
\left\vert c_{11}\right\vert  &\gtrsim &\alpha _{0}\left( 1-13.6\tau
^{2}\right) , \\
\left\vert c_{12}\right\vert  &\lesssim &0.8\alpha _{0}\tau ^{2}, \\
\left\vert c_{21}\right\vert  &\lesssim &0.8\alpha _{0}\tau , \\
\left\vert c_{22}\right\vert  &\lesssim &\alpha _{0}\tau,
\end{eqnarray}
\end{subequations}
where the parameter $\tau = h r_{0}$.
We see that under  the  condition  $\tau  \ll  1$,  the  absolute  value of the
$c_{11}$ component is  much  larger than those of the other three components of
the radial wave function.

When the distance from the core  of  the  soliton is sufficiently large, we can
neglect the fermion-soliton interaction.
In this case, the general solution to system (\ref{III:7}) is written as
\begin{subequations}                                                \label{B:3}
\begin{eqnarray}
c_{11}\left( r\right)  &\sim &C_{1}J_{0}\left( pr\right) +C_{2}Y_{0}\left(
pr\right) , \\
c_{12}\left( r\right)  &\sim &-\sqrt{\frac{\varepsilon -h}{\varepsilon +h}}
\left( C_{3}J_{1}\left( pr\right) +C_{4}Y_{1}\left( pr\right) \right) , \\
c_{21}\left( r\right)  &\sim &-\sqrt{\frac{\varepsilon -h}{\varepsilon +h}}
\left( C_{1}J_{1}\left( pr\right) +C_{2}Y_{1}\left( pr\right) \right) , \\
c_{22}\left( r\right)  &\sim &C_{3}J_{2}\left( pr\right) +C_{4}Y_{2}\left(
pr\right),
\end{eqnarray}
\end{subequations}
where $C_{1}$ -- $C_{4}$ are constant coefficients, $J_{\nu}$ and $Y_{\nu}$ are
the Bessel and Neumann functions of corresponding orders, respectively.
Using standard methods  from the theory of scattering \cite{LandauIII, Taylor},
we obtain the expressions  for the  partial elements of the $S$-matrix in terms
of the coefficients $C_{1}$ -- $C_{4}$
\begin{subequations}                                                \label{B:4}
\begin{eqnarray}
S_{1/2,1/2}^{\left( +1\right) } &=&\frac{C_{1}-iC_{2}}{C_{1}+iC_{2}}, \\
S_{-1/2,1/2}^{\left( +1\right) } &=&\frac{2p}{\varepsilon -h}\frac{C_{3}}
{C_{1}+iC_{2}}.
\end{eqnarray}
\end{subequations}
To express the coefficients $C_{1}$ -- $C_{4}$ in terms of physical parameters,
we join  solutions  (\ref{B:1})  and  (\ref{B:3})  at $r = \kappa r_{0}$, where
$\kappa$ is a positive coefficient greater than one.
The  resulting  expressions  for  $C_{1}$ -- $C_{4}$  are  too  lengthy  to  be
presented here.
Expanding these expressions in $p$, holding lower-order terms, and substituting
the resulting expressions  in  Eq.~(\ref{B:4}),  we  obtain  the partial matrix
elements:
\begin{equation}
S_{1/2,1/2}^{\left( +1\right) }=\frac{A+\pi ^{-1}\Gamma \ln \left[ pr_{0}
\right]+i\Gamma/2}{A+\pi^{-1}\Gamma \ln\left[pr_{0}\right] -
i\Gamma /2},                                                        \label{B:5}
\end{equation}
and
\begin{equation}
S_{-1/2,1/2}^{\left( +1\right) }=\frac{-i\tau \Gamma }{A+\pi ^{-1}\Gamma \ln
\left[ pr_{0}\right] -i\Gamma /2}                                   \label{B:6}
\end{equation}
where
\begin{eqnarray}
A &=&1+\kappa ^{2}-\frac{1}{6}\tau ^{2}\left( \left(12 \gamma + \pi
^{2}\right) \kappa^{2}+\pi ^{2}\right)                             \label{B:7}
\\
&&+\frac{1}{2}\tau ^{2}\left\{ 2\kappa ^{2}\ln \left[ 4\left( \kappa
^{2}+1\right) \kappa^{-2}\right] \right. \nonumber \\
&&\left. +\left\{ 4\left(\kappa^{2}+1\right) \left( \gamma + \ln \left[
2^{-1}\left( \kappa ^{2}+1\right)^{1/4} \right] \right) +2\right\} \right.
\nonumber \\
&&\times \left. \ln\left[\kappa^{2}+1\right]\right\}+\tau^{2}\left(\kappa
^{2}+1\right) \text{Li}_{2}\left[ \left( \kappa^{2}+1\right)^{-1}
\right], \nonumber
\end{eqnarray}
\begin{equation}
\Gamma=2\pi \tau^{2}\left(\left(\kappa^{2}+1\right) \ln \left[\kappa^{2}+1
\right] - \kappa ^{2}\right),                                       \label{B:8}
\end{equation}
and $\gamma$ is the Euler-Mascheroni constant.

\begin{figure}[tbp]
\includegraphics[width=0.5\textwidth]{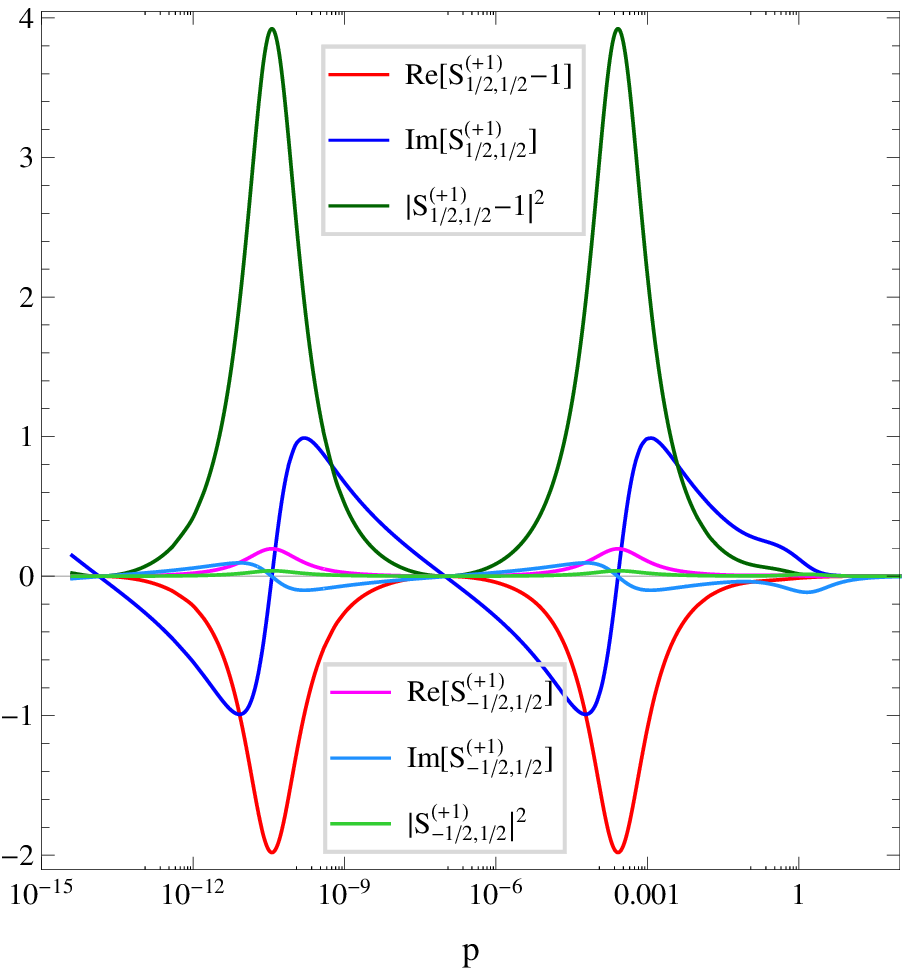}
\caption{\label{fig8}     Dependence of $\text{Re}[S_{1/2,1/2}^{\left(+1\right)
}-1]$, $\text{Im}[S_{1/2,1/2}^{\left(+1\right)}]$, $\vert S_{1/2,1/2}^{\left(+1
\right)}\vert^{2}$, $\text{Re}[S_{-1/2,1/2}^{\left(+1\right)}]$, $\text{Im}[S_{
-1/2,1/2}^{\left(+1\right)}]$, and $\vert S_{-1/2,1/2}^{\left(+1\right)}\vert^{
2}$ on the fermion momentum $p$.}
\end{figure}

\begin{figure}[tbp]
\includegraphics[width=0.5\textwidth]{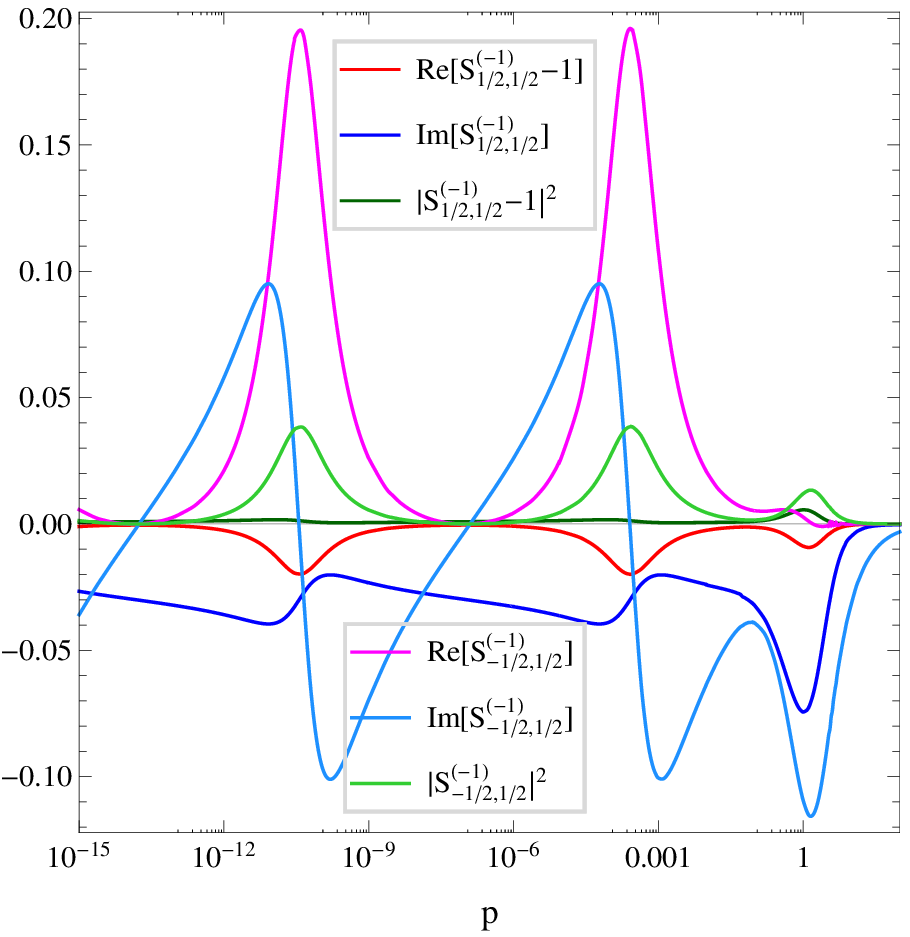}
\caption{\label{fig9}     Dependence of $\text{Re}[S_{1/2,1/2}^{\left(-1\right)
}-1]$, $\text{Im}[S_{1/2,1/2}^{\left(-1\right)}]$, $\vert S_{1/2,1/2}^{\left(-1
\right)}\vert^{2}$, $\text{Re}[S_{-1/2,1/2}^{\left(-1\right)}]$, $\text{Im}[S_{
-1/2,1/2}^{\left(-1\right)}]$, and $\vert S_{-1/2,1/2}^{\left(-1\right)}\vert^{
2}$ on the fermion momentum $p$.}
\end{figure}

Using   Eqs.~(\ref{B:5}),   (\ref{B:6}),  and  (\ref{III:17}),  we  obtain  the
expressions for the squared magnitudes of the partial amplitudes $f_{1/2,1/2}^{
\left(+1\right)}$ and $f_{-1/2,1/2}^{\left(+1\right)}$:
\begin{equation}
\left\vert f_{1/2,1/2}^{\left( +1\right) }\right\vert ^{2}=\frac{2}{\pi p}
\frac{\left( \Gamma /2\right) ^{2}}{\left( A+\pi ^{-1}\Gamma \ln \left[
pr_{0}\right] \right) ^{2}+\left( \Gamma /2\right) ^{2}},           \label{B:9}
\end{equation}
and
\begin{equation}
\left\vert f_{-1/2,1/2}^{\left( +1\right) }\right\vert ^{2}=\frac{1}{2\pi p}
\frac{\Gamma _{r}\Gamma }{\left( A+\pi ^{-1}\Gamma \ln \left[ pr_{0}\right]
\right) ^{2}+\left( \Gamma /2\right) ^{2}},                        \label{B:10}
\end{equation}
where
\begin{equation}
\Gamma_{r} = \tau^{2}\Gamma.                                       \label{B:11}
\end{equation}
We can see that the squared magnitudes of the partial amplitudes $f_{1/2,1/2}^{
\left(+1\right)}$ and $f_{-1/2,1/2}^{\left(+1\right)}$, considered as functions
of  the  logarithmic  variable $\pi^{-1} \Gamma \ln \left[p r_{0}\right]$, show
resonance behavior of the  Breit-Wigner  type  with  total decay width equal to
$\Gamma$.
It follows that on  the  logarithmic scale, the  squared  magnitudes $\vert S_{
I_{3}^{\prime },I_{3}}^{\left(+1\right)}-\delta_{I_{3}^{\prime},I_{3}}\vert^{2}
= 2\pi p\vert f_{I_{3}^{\prime},I_{3}}^{\left( +1 \right)}\vert^{2}$ are of the
symmetric  Breit-Wigner  form   in   the   resonance   region,   as  it  is  in
Fig.~\ref{fig8}.
At the same  time, the  resonance  peaks  of  $\vert S_{I_{3}^{\prime},I_{3}}^{
\left(+ 1 \right)} - \delta_{I_{3}^{\prime}, I_{3}} \vert^{2}$  have a strongly
asymmetric form in the linear scale.
In particular, the resonance peaks  have  maxima  at $p_{\max} = r_{0}^{-1}\exp
\left(- \pi A/\Gamma \right)$  and  reach  half of the maxima at $p_{\pm 1/2} =
r_{0}^{-1}\exp \left( -\pi A/\Gamma \pm \pi /2\right)$.
The asymmetry of the resonance peaks is reflected in the relations
\begin{subequations}
\begin{equation}
\frac{p_{+1/2}}{p_{\max }}=\frac{p_{\max }}{p_{-1/2}} =
\exp \left( \pi/2\right) \approx 4.81048,
\end{equation}
and
\begin{equation}
\frac{p_{\max }-p_{-1/2}}{p_{+1/2}-p_{\max }}=\exp \left( -\pi /2\right)
\approx 0.20788.
\end{equation}
\end{subequations}
The position of the maxima of the  resonance peaks depends exponentially on the
ratio $A/\Gamma$, which, in turn, depends on the parameters $\tau$ and $\kappa$.
This results in extremely low values of $p_{\max}$ when $\tau \lesssim 0.1$ and
$1 \lesssim \kappa \lesssim 10^{2}$  in  accordance  with  Figs.~\ref{fig8} and
\ref{fig9}.

The existence of the resonance peaks  and  their logarithmic character  are due
to the term $\ln\left[p r_{0}\right]$ in Eqs.~(\ref{B:5}) and (\ref{B:6}).
In turn, this term results  from  the  leading  term  of  the  expansion of the
Neumann function  $Y_{0}(pr)$  in  the  neighborhood  of  zero: $Y_{0}\left(p r
\right) = 2\pi^{-1}\left(\gamma-\ln \left( 2\right)+\ln\left[p r\right] \right)
+ O(\left(pr\right)^{2}\ln\left[p r\right])$.
Only the Neumann  function  $Y_{0}(pr)$  has  the  leading  asymptotic behavior
$\propto \ln[p r]$ as $p r \rightarrow 0$, whereas  the other Neumann functions
$Y_{l>0}(pr) \propto \left( pr\right) ^{-l} $ in this limit.
Therefore, the Neumann functions $Y_{l>0}(pr)$ cannot lead  to  resonance peaks
of the logarithmic type.

It follows from  Eq.~(\ref{B:5}) that  the  squared magnitude of $S_{1/2,1/2}^{
\left(+1\right)}$ is equal to one.
Therefore, the unitarity condition $\vert S_{1/2,1/2}^{\left(+1\right)}\vert^{2}
+ \vert S_{-1/2,1/2}^{\left( +1\right) } \vert^{2} = 1$ is not satisfied in the
used approach.
Eq.~(\ref{B:6}) tells us that the squared magnitude  of $S_{-1/2,1/2}^{\left(+1
\right)}$ does not exceed the value of $4\tau^{2} = \left(2h r_{0}\right)^{2}$,
which  is  equal  to  $0.04$  for  the  values  of  $h$  and  $r_{0}$  used  in
Sec.~\ref{sec:V}.
It follows that the  violation  of  the  unitarity  is small when the parameter
$\tau = h r_{0}$ is much smaller than one.

The partial matrix  elements  (\ref{B:5})  and  (\ref{B:6})  have a simple pole
located at $p_{\text{pol}} = i r_{0}^{-1}\exp\left(-\pi A/\Gamma\right) = i p_{
\max}$. %the imaginary  fermion  momentum
According  to  the  theory of scattering \cite{LandauIII, Taylor}, a pole of an
elastic partial element of  the  $S$-matrix  located  at  a  positive imaginary
momentum corresponds to a bound state.
In our case,  however,  the  presence  of  the  pole is only an artifact of the
used approach.
Indeed, this pole exists for arbitrary small values  of  $h$ and $r_{0}$, which
is impossible for bound states.
We also failed to find fermion  bound  states  in  the partial channels with $m
= \pm 1$ using numerical methods.

The obvious drawback of  the used approach  is that it is unable to explain the
periodic structure  of the resonance peaks in Figs.~\ref{fig8} and  \ref{fig9}.
This approach, however, can  explain  the  location  of  the resonance peaks at
extremely low fermion momenta and  the  Breit-Wigner form of these peaks on the
logarithmic scale.

\bibliography{article}
\clearpage

\clearpage

\clearpage

\clearpage

\clearpage

\end{document}